\documentclass[utf8,prologue,dvipsnames]{FrontiersinHarvard} 

\usepackage{url,hyperref,microtype,subcaption}
\usepackage[onehalfspacing]{setspace}

\usepackage{multirow}
\usepackage[english]{babel}
\usepackage[autostyle=true,english=american]{csquotes}
\usepackage{acronym}
\graphicspath{{figures/}}
\usepackage{booktabs}

\newacro{AI}[AI]{Artificial Intelligence}
\newacro{UI}[UI]{user interface}
\newacro{GUI}[GUI]{graphical user interface}
\newacro{TLX}[TLX]{NASA-Task Load Index}
\newacro{RTLX}[Raw-TLX]{NASA Raw-Task Load Index}
\newacro{ER}[ER]{error rate}
\newacro{TCT}[TCT]{task completion time}
\newacro{HCI}[HCI]{Human-Computer Interaction}
\newacro{UX}[UX]{user experience}
\newacro{HFE}[HFE]{Human Factors and Ergonomics}
\newacro{cuDNN}[cuDNN]{CUDA Deep Neural Network library}
\newacro{RMSE}[RMSE]{root mean squared error}
\newacro{HMD}[HMD]{Head-Mounted Display}
\newacro{RF}[RF]{Random Forest}
\newacro{GP}[GP]{Gaussian process, long-plural = Gaussian processes}
\newacro{KNN}[\textit{k}NN]{\textit{k}-nearest neighbor}
\newacro{NN}[NN]{Neural Network}
\newacro{DNN}[DNN]{ Deep Neural Network}
\newacro{CNN}[CNN]{Convolutional Neural Network}
\newacro{FCL}[FCL]{fully connected layer}
\newacro{BoD}[BoD]{Back-of-Device}
\newacro{FOV}[FoV]{field of view}
\newacro{RW}[RW]{Real World}
\newacro{IFRC}[IFRC]{index finger ray cast}
\newacro{FRC}[FRC]{forearm ray cast}
\newacro{EFRC}[EFRC]{eye-finger ray cast}
\newacro{HRC}[HRC]{Human-Robot Collaboration}
\newacro{HRI}[HRI]{Human-Robot Interaction}
\newacro{6DOF}[6DOF]{six-degree-of-freedom}
\newacro{LOOCV}[LOOCV]{leave-one-out cross-validation}
\newacro{CV}[CV]{cross-validation}
\newacro{RM}[RM]{repeated measure}
\newacro{ANOVA}[ANOVA]{analysis of variance}
\newacro{RMANOVA}[RM-ANOVA]{repeated measures analysis of variance}
\newacro{AGATe}[AGATe]{AGreement Analysis Toolkit}
\newacro{GHoST}[GHoST]{Gesture Heatmap Toolkit Gesture Heatmaps Toolkit}
\newacro{GREAT}[GREAT]{Gesture Relative Accuracy Toolkit}
\newacro{GRT}[GRT]{Gesture Recognition Toolkit}
\newacro{DTW}[DTW]{Dynamic Time Warping}
\newacro{LHRD}[LHRD]{large high resolution display}
\newacro{GEQ}[GEQ]{Game Experience Questionnaire}
\newacro{SPGQ}[SPGQ]{Social Presence Gaming Questionnaire}
\newacro{JND}[JND]{just-noticeable difference}
\newacro{SUS}[SUS]{system usability scale}
\newacro{CSCW}[CSCW]{computer-supported cooperative work}
\newacro{CAD}[CAD]{computer-aided design}
\newacro{MR}[MR]{Mixed Reality}
\newacro{CVE}[CVE]{Collaborative Virtual Environment}
\newacro{AR}[AR]{Augmented Reality}
\newacro{AV}[AV]{Augmented Virtuality}
\newacro{VR}[VR]{Virtual Reality}
\newacro{PRISMA}[PRISMA]{Preferred Reporting Items for Systematic Reviews}
\newacro{PRISMA-Scope}[PRISMA-ScR]{Meta-Analyses Extension for Scoping Reviews}
\newacro{TF-IDF}[TF-IDF]{Term Frequency-Inverse Document Frequency}
\newacro{TF}[TF]{Term Frequency}
\newacro{AVs}[AVs]{Automated Vehicles}
\newacro{eHMIs}[eHMIs]{external Human-machine interfaces}
\newacro{SAR}[SAR]{Spatial Augmented Reality}
\newacro{IFR}[IFR]{International Federation of Robotics}
\newacro{ADLs}[ADLs]{Activities of Daily Living}
\newacro{LED}[LED]{Light-Emitting Diode}
\newacro{DoF}[DoF]{Degree-of-Freedom} \newacroplural{DoF}[DoFs]{Degrees-of-Freedom}
\newacro{HHC}[HHC]{Human-Human Collaboration}
\newacro{IDF}[IDF]{Inverse Document Frequency}
\newacro{QUEAD}[QUEAD]{Questionnaire for the Evaluation of Physical Assistive Devices}
\newacro{TiA}[TiA]{Trust in Automation Questionnaire}
\newacro{TOR}[TOR]{Take-Over-Request}
\newacro{ADMC}[ADMC]{Adaptive DoF Mapping Controls}

\newacro{MT}[MT]{Machine Translation}
\newacro{AI}[AI]{Artificial Intelligence}
\newacro{LLM}[LLM]{Large Language Model}
\newacro{NLP}[NLP]{Natural Language Processing}
\newacro{NRC}[NRC]{U.S. National Research Council}
\newacro{ALPAC}[ALPAC]{Automatic Language Processing Advisory Committee}
\newacro{SUR}[SUR]{Speech Understanding Research}
\newacro{NN}[NN]{neural network}
\newacro{DARPA}[DARPA]{Defense Advanced Research Projects Agency}
\newacro{MAS}[MAS]{Multi-agent System}
\newacro{LSTM}[LSTM]{Long Short-Term Memory}
\newacro{BP}[BP]{backpropagation}
\newacro{WW2}[WW2]{World War II}
\newacro{MIT}[MIT]{Massachusetts Institute of Technology}
\newacro{IPL}[IPL]{Information Processing Language}
\newacro{GPT}[GPT]{Generative pre-trained transformer}
\newacro{BERT}[BERT]{Bidirectional Encoder Representations from Transformers}
\newacro{NLUI}[NLUI]{Natural-language user interface}
\newacro{NLI}[NLI]{Natural language interface}
\newacro{CUI}[CUI]{Conversational user interface}
\newacro{VUI}[VUI]{Voice user interface}
\newacro{MRS}[MRS]{Multi-robot system}
\newacro{AAL}[AAL]{Ambient Assisted Living}
\newacro{CNN}[CNN]{convolutional neural network}
\newacro{RNN}[RNN]{recurrent neural network}
\newacro{SDS}[SDS]{Spoken Dialogue System}
\newacro{VPA}[VPA]{Virtual Personal Assistant}
\newacro{IPA}[IPA]{Intelligent Personal Assistant}
\newacro{SAR}[SAR]{Socially Assistive Robot}
\newacro{SASSI}[SASSI]{Subjective Assessment of Speech System Interfaces}
\newacro{RQ}[RQ]{research question}
\newacro{JSON}[JSON]{JavaScript Object Notation}
\newacro{ROS}[ROS]{Robot Operating System}
\newacro{API}[API]{Application Programming Interface}

\def\keyFont{\fontsize{8}{11}\helveticabold }
\def\firstAuthorLast{Y. Lakhnati {et~al.}} 
\def\Authors{Younes Lakhnati\,$^{1,*}$, Max Pascher\,$^{1,2,*}$, and Jens Gerken\,$^{1,*}$}
\def\Address{$^{1}$TU Dortmund University, Inclusive Human-Robot-Interaction, Dortmund, NW, Germany \\
$^{2}$University of Duisburg-Essen, Human-Computer Interaction, Essen, NW, Germany}
\def\corrAuthor{Younes Lakhnati, Max Pascher, Jens Gerken}
\def\corrEmail{TU Dortmund University, Emil-Figge-Straße 50, 44227 Dortmund, NW, Germany, \{firstname\}.\{lastname\}@udo.edu}

\newcommand\change[1]{{#1}}

\begin{document}
\onecolumn
\firstpage{1}

\title[GPT for Variable Autonomy]{Exploring a GPT-based Large Language Model for Variable Autonomy in a VR-based Human-Robot Teaming Simulation}

\author[\firstAuthorLast ]{\Authors} 
\address{} 
\correspondance{} 

\extraAuth{} 

\maketitle
\begin{abstract}
\section{}
In a rapidly evolving digital landscape autonomous tools and robots are becoming commonplace. Recognizing the significance of this development, this paper explores the integration of \acp{LLM} like \textit{\ac{GPT}} into human-robot teaming environments to facilitate variable autonomy through the means of verbal human-robot communication.
In this paper, we introduce a novel \change{simulation} framework for such a \ac{GPT}-powered multi-robot testbed environment, based on a Unity \ac{VR} setting. This system allows users to interact with \change{simulated robot agents} through natural language, each powered by individual \ac{GPT} cores. By means of OpenAI's function calling, we bridge the gap between unstructured natural language input and \change{structured} robot actions.
A user study with 12 participants explores the effectiveness of \ac{GPT}-4 and, more importantly, user strategies when being given the opportunity to converse in natural language within a simulated multi-robot environment. Our findings suggest that users may have preconceived expectations on how to converse with robots and seldom try to explore the actual language and cognitive capabilities of their simulated robot collaborators. Still, those users who did explore where able to benefit from a much more natural flow of communication and human-like back-and-forth. We provide a set of lessons learned for future research and technical implementations of similar systems. 

\tiny
 \keyFont{ \section{Keywords:} assistive robots, virtual reality, evaluation, shared control, variable autonomy} 
\end{abstract}

\section{Introduction}
\label{sec:introduction}
\change{Over the past 10 to 15 years, we have seen a significant shift from machines being just useful tools towards machines as partners in more complex human like collaboration tasks~\citep{Grudin.2017}. Still, true peer-to-peer interaction between humans and machines, such as robots, still has a long way to go to resemble inter-human interaction and collaboration.} 
\change{Importantly}, for humans to cooperate seamlessly with autonomous tools, such as robots, a delicate balance must be struck between both a user's need for autonomy~\change{\citep{deci2012self}} and the robot's capabilities and level of autonomous behavior. 
\change{In this context, concepts such as variable autonomy and shared control have been vital research areas.}
Shared control, in particular, aims to find optimal configurations of how control can be shared between an autonomous machine such as a robot and human so that the overall system is as effective and efficient as possible~\citep{ERDOGAN2017282}. It requires, at its core, an effective way for communication between parties to negotiate and adapt the sharing of control \citep{flemisch2019joining, Abbink2018,Pascher.2023robotMotionIntent}. While such communication between human and machines can rely on a variety of input and output modalities, spoken language, as a vehicle of thought, has played an essential role for human-human communication and collaboration~\citep[pp.~17--18]{clark1996using}. The exploration of natural language in fields like human-machine cooperation and AI has a long and storied history. \ac{NLP} has been an incredibly significant branch of robotics and \ac{AI} research since its infancy, spearheaded by foundational works like Winograd's \emph{SHRDLU}~\citep{Winograd.1972} and Nilsson et al.'s \emph{Shakey the Robot}~\citep{nilsson1984shakey}. 

Conceptually, however, for language to be an effective means for collaboration, research has suggested the overarching concept of \emph{Joint Actions}~\citep[p.~59]{clark1996using}, which has become a popular approach to understanding human-human cooperation \citep{sebanz2006joint, garrod2004conversation, vesper2010minimal}. This concept has the potential to be extrapolated to the human-and-machine side of cooperation, as suggested by \citet[p.~2]{flemisch2019joining}. \change{Since this perspective suggests that joint activity and communication are inseparable and that communication requires the use of a common language, the progress in this field is heavily tied to the machine's abilities to understand and communicate in natural language. It is within this context that the field of \ac{NLP} has witnessed significant breakthroughs as of late. These technological advancements, epitomized by the development of the Transformer architecture by \citeauthor{vaswani2017attention}, have lead to the emergence of \acp{LLM} such as \ac{GPT} and \textit{\ac{BERT}}~\citep[p.~6]{radford2018improving} \citep[p.~6]{devlin2018bert}. These have demonstrated competences in fields such as translation \citep[p.~6--7]{brown2020language}, question-answering \citep{Nakano.17.12.2021}, creative writing \citep{Swanson.2021} and even medical diagnostics \citep{hirosawa2023diagnostic,nori2023capabilities}.}


\change{We believe that the capability of \acp{LLM} to interpret natural language inputs and dynamically respond to them highlights their potential utility in the realm of shared control and variable autonomy. These systems, by design, can engage in a form of interaction that is intuitive to humans, leveraging our natural propensity for verbal communication as seen in cooperative human-to-human interactions. The ability of \acp{LLM} to generate responses that convincingly resemble human dialogue makes them well-suited for mediating communication between humans and autonomous agents, providing a common medium through which shared control and variable autonomy can be negotiated and dynamically adjusted. The linguistic bridge that \acp{LLM} provide could allow for the articulation of intentions, feedback, and commands between humans and machines, thereby facilitating the implementation of shared control and variable autonomy in a way that is intuitive and aligned with human cognitive processes.}
Of course, there are certain inherent challenges in play as well. The use cases where \acp{LLM} excel do not require to map spoken commands to a very specific output that is bound by the physical capabilities of an autonomous agent. Still, existing work has shown that available \acp{LLM} such as \ac{GPT} might already allow the communication and coordination based on natural language with a robot~\citep{koubaa2023rosgpt}.
This paper aims to explore this up and coming research area both from a feasibility and a user perspective. 

 To this end, we have designed an \ac{LLM}-powered multi-agent system as part of a \ac{VR}-based \change{simulation} framework. In the virtual game world, a user can fulfill simple tasks solely by conversing with three \change{simulated robot agents}, which are controlled by their own \ac{GPT} cores. Utilizing the prototype, a user study spanning 12 participants was conducted to investigate the ways in which humans engaged with the \change{simulated robot agents}, how natural it felt to them and how well \ac{GPT} was suited for this use case.  \change{A key area of interest in our exploration was understanding user reactions and actions, especially when the \ac{GPT}-based simulated robot agents did not behave as anticipated, highlighting the nuances of human-agent interaction.}
Consequently, the contribution of this paper consists of 

\begin{itemize}
    \item The design and architecture of a Unity-based framework for a voice-controlled multi-agent system in VR, with which the interaction and control dynamics between a single user and a scalable amount of \ac{LLM}-based \change{simulated robot agents} can be explored\footnote{We plan to make the framework available as Open-Source on GitHub upon acceptance and are happy to provide it to reviewers on request in the meantime}.
    \item Based on our study, an improved understanding of user's strategies and behaviors when being confronted with an \ac{LLM}-based \change{simulated multi-robot environment}.
    \item A set of lessons learned on the feasibility and practicability of adapting \ac{GPT}-based \acp{LLM} to interact with \change{simulated} multi-robot agents. 
    \change{This includes reflections on the naturalness of such interactions and the adaptability of \ac{GPT} and its users in unforeseen scenarios, as well as possible improvements to the framework that could inspire other similar systems.}
\end{itemize}

\change{
Adding to the contributions listed, it is important to note that our Unity simulation framework utilizes \emph{simulated} robot agents, deliberately focusing on the communication interface between humans and \ac{LLM}-powered agents rather than the mechanical intricacies of physical robots and actions. This approach enables a detailed examination of human-agent communication, relevant both to hypothetical real-world applications and to theoretical explorations of \ac{LLM} capabilities in robotics. 

While our approach is exploratory in nature, we emphasize that exploratory research methods are well-established in the literature as a means to investigate areas where there is limited pre-existing knowledge or to provide a comprehensive understanding of a new or complex issue. This is also applicable in the field of \ac{HRI}, as demonstrated by the review from \citeauthor{veling2021qualitative} of similar qualitative research approaches. In the same vein, our simulation framework is an exploratory tool, specifically aimed at other researchers outside the field of \ac{AI} or traditional robotics who may want to further investigate the interaction dynamics between human users and simulated \ac{LLM}-based robot agents.}

\section{Background}
\label{sec:background}
\subsection{Shared Control \& Variable Autonomy}
\change{While research in autonomous robotics is a foundational and ongoing area of inquiry within the field~\citep{Canal.2016, lauretti2017learning, Gallenberger.2019, rakhimkul2019autonomous}, there are several indications that relying solely on an autonomous robot does not necessarily fulfill basic psychological needs such as self-autonomy and competence.}
For example, \citet{Kim.2012} reported that the sporadic and supervisory nature of working with an autonomous robot results in the users experiencing them as \enquote{one more external agent that does their work for them}. These findings are corroborated by \citet{Pollak.2020autonomystress} who reported that manual control compared to autonomous behavior led to significantly reduced symptoms of stress and a lower heart rate. Similarly, \citet{zlotowski2017can} found that autonomous robots generally evoke a more negative attitude than non-autonomous robots and are experienced as more threatening. \citet{latikka2021attitudes} reported, that in particular in the workplace, users preferred to see robots as equipment and as a result preferred non-autonomous robots. In addition \citet{Park.2020} showed that fully autonomous robots may discourage people with motor impairments from using their remaining physical capabilities.

To counter that, shared-control (or traded-control) systems aim to strike a balance between autonomous robot behavior and manual user control \citep{ERDOGAN2017282, Pascher.2023inTimeAndSpace,Pascher.2024adaptix}. Due to the lack of clear definitions of these terms, \citet{Abbink2018} introduced a topology of shared control systems and axioms for the design and evaluation thereof, unifying varying shared control concepts and definitions under one common framework. \citet{flemisch2019joining} later expanded upon \citeauthor{Abbink2018}'s framework and explored shared-control in the context of human-robot cooperation, emphasizing the importance of cooperation on multiple layers of interaction \change{and control}, such as a strategical, tactical and operational layer.

\citet{Abbink2018} argue, that robot functionality, and as a result \change{user} preferences and abilities, are situational and depend on context. As a result, concepts such as variable autonomy or adjustable autonomy have emerged, which build upon the principle of shared control by adding a dynamic and situated component \citep{Chiou.2023vat, Bustamante2021}. 
\change{These concepts often enable the user to adjust their level of involvement or control in a task, typically through user interface elements like buttons. However, there has been a longstanding research interest in using language as a means of collaboration with machines.}

\subsection{Natural Language Interface systems and Large Language Models}
The domain of \acp{NLI} represents an important reference point for our research. Historical works such as \textit{SHRDLU}~\citep{Winograd.1972} and \textit{LUNAR system}~\citep{Woods.1973} are early examples of natural language being used to control complex systems. Both Woods and Winograd identified that the biggest challenges stemmed from semantic barriers that would need to be overcome by advancements in the field of \ac{NLI} and \ac{NLP}. While still battling with the challenges of human linguistics, commercial product developments in the form of \acp{VPA} or \acp{IPA}, such as Microsoft's Cortana, Apple's Siri, Amazon Alexa, Google Assistant, and so on were able to create a public awareness, establish a user base and provide a better understanding of the potential of such systems \citep{kepuska2018next}. Related approaches were also introduced into the field of robotics -- here \citet{liu2019review} offer an extensive overview related to \ac{NLI}-controlled robots.
While these approaches were certainly impressive, they still lacked a fundamental understanding that transcends their confined domains.

Following \citet{vaswani2017attention}'s breakthrough work on self-attention, the development of transformer models has reshaped the landscape of \ac{AI} as well as \ac{NLP} and gave rise to \acp{LLM}: AI systems, which are trained on massive amounts of textual data using deep learning techniques. Building upon this concept, \ac{BERT}, revolutionized many \ac{NLP} tasks by training bidirectionally, meaning it considers both the left and the right context in all layers, capturing information from the entire passage \citep{devlin2018bert}. This made BERT particularly suited for tasks that require understanding context. On the other hand, \ac{GPT} is a unidirectional model (every token can only attend to previous tokens) trained to predict the next word in a sequence \citep{radford2018improving}. Despite this, it excels in various \ac{NLP} tasks by leveraging its transformer architecture and a massive amount of data for training. \ac{GPT} is especially noteworthy for its capability to generate coherent and contextually relevant text over long passages. Both, \ac{BERT} and \ac{GPT} have not only been released for research collaboration, but have been integrated into commercial products and provide \ac{API} access for software developers. Noteworthy, there is a sheer infinite number of further \acp{LLM} and several significant extensions to these existing approaches \citep{yang2019xlnet, brown2020language, chen2021evaluating, neelakantan2022text, stiennon2020learning, ouyang2022training}. Thanks to the generality of language, these models have been applied in various applications, ranging from search engines and chat bots \citep{kelly2023bing}, general problem-solving and knowledge extraction \citep{petroni2019language}, medical diagnosis \citep{hirosawa2023diagnostic, nori2023capabilities, waisberg2023gpt, shea2023use, rao2023evaluating}, education \citep{tack2022ai, firat2023chat, ausat2023can}, law \citep{cyphert2021human, perlman2022implications, trozze2023large} and robotics, as we will discuss in the next section.

\subsection{LLMs and Robots}
\acp{LLM} have opened up new possibilities in the field of robotics and human-robot teaming, most apparently for social robots and \acp{SAR} \citep{lee2023developing, alessa2023towards, kahambing2023chatgpt, lekova2023system, irfan2023between}. 
\change{However, the conversational capabilities of \acp{LLM} extend beyond mere social interactions; their proficiency in handling a diverse range of textual inputs - without the need for rigidly predefined formats - marks a significant advancement for applications such as speech-controlled robotics, which have historically faced challenges with processing unstructured input.}
Recent research indicates that \acp{LLM} work well in this environment and that users generally prefer unstructured modes of communication in comparison to structured ones \citep[p.~10]{kodur2023structured}. 

\change{One core challenge lies in the interpretation of unstructured verbal input and its translation into structured robot actions, a process that is crucial for effective human-robot interaction. This involves not only the interpretation and translation of verbal commands into actionable tasks but also the nuanced process of language grounding. Language grounding specifically relates to the ability of the system to connect linguistic symbols with their real-world referents, ensuring that the robot's understanding of commands is deeply rooted in the physical context it operates within.}

\citet{trott2015natural}, while not yet being able to rely on a \ac{LLM}, presented a promising architectural layout for verbal communication with multiple \change{simulated robot agents}. They introduced a boss-agent, which routes the communication between multiple \change{simulated robot agents} and the user via N-tuples, which is translated into a pattern response. While this strategy was limited by the need for an explicit grammar and lexicon, similar to the grammar-based method by \citet{Misra2015}, we adapted Trott et al.'s approach for our work, as we will discuss in Section \ref{sec:framework}. \citet{Arumugam.2019} analyzed the issue of language grounding more closely and proposed a strategy to achieve a high \enquote{accuracy, efficiency and generalization through the construction of simple, semantic goal representations within Markov decision processes}. \citet{ahn2022can} propose a method called \enquote{SayCan}, that integrates the capabilities of \acp{LLM} with robotic affordances (learned mappings that quantify the likelihood of a robot successfully executing specific actions given its current state) to better interpret and execute high-level textual instructions. Their approach seeks to bridge the gap between the knowledge embedded within \acp{LLM} and the practical constraints of physical environments. Similar to our work, \citet{koubaa2023gpt} integrated a \ac{GPT} model and a parser to operate a robot -- in this case a real robot - through \ac{ROS}. They employed an ontology to map unstructured language input into structured robotic instructions, but encountered issues with the model unexpectedly straying from the ontology at times. We initially chose the same approach and encountered similar issues of \ac{GPT} not adhering to the ontology.

More recently, \citeauthor{handler2023balancing} introduced a multi-dimensional taxonomy, specifically designed to evaluate and categorize the way autonomous \ac{LLM}-powered multi-agent systems manage the relationship between autonomy and alignment \cite{handler2023balancing}. This analysis covers various architectural perspectives, focusing on their intrinsic characteristics and operational mechanisms. Händler's approach is motivated by the observation that existing taxonomies fall short to categorize and understand the complexity of these new systems.

Aiming to understand how users interact with machines using language, the work of Porcheron and colleagues is notable. Starting with \acp{IPA}, \citeauthor{porcheron2017animals} examined the integration of IPAs within human conversation, focusing on linguistic aspects. They explore how IPAs such as Siri and Alexa influence conversational dynamics and shed light on the linguistic interplay between human users and AI agents \citep{porcheron2017animals}. Expanding on this theme, \citeauthor{fuentes2023roboclean} explore linguistic challenges faced in human-robot interactions in low-resource environments \citep{fuentes2023roboclean}. Their research emphasized the importance of robots possessing a deep contextual understanding to accurately interpret user instructions. They identified natural language challenges such as referential expression resolution and the dynamic nature of language, which can pose hurdles in human-robot communication. Their findings accentuate the need for improved linguistic reasoning in robots, especially in specialized environments where context plays a key role.

\subsection{Take-away}
In our work, we add to this emerging area of research by addressing two perspectives in particular. First, we realized that while the utilization and integration of \acp{LLM} has become much easier, prototyping \ac{LLM}-powered \change{simulated robot agents} is not straightforward, which limits our ability to explore the inherent trade-offs and limitations. Therefore, we explored and developed the integration of a \ac{GPT}-4 model within a Unity-based simulation environment, relying on the just introduced function call capabilities to facilitate the mapping of unstructured speech input to structured robot actions. 
\change{While there has been research indicating that users adapt their way of speaking when interacting with robots \citep{fussell2008people, pelikan2016nao}, research exploring similar adaptations in conversations with \ac{LLM}-based agents is still emerging. Given the increasing sophistication of \acp{LLM} and their potential for more nuanced understanding and generation of natural language, investigating how human speech patterns adapt and how \ac{LLM} answers influence these interactions could yield valuable insights into human-\ac{AI} communication dynamics.}
\change{This is particularly important given the unpredictable behavior of \acp{LLM}, which necessitates practical exploration in implementations involving \ac{LLM}-powered simulated robot agents. While studies such as \citet{fuentes2023roboclean} use Wizard-of-Oz techniques to mostly simulate a perfectly functioning \ac{AI}, we believe that human-robot conversation may be most interesting in situations where there is miscommunication. To that end, we propose that studying actual \ac{LLM} implementations where \ac{LLM}-powered agents have to act, cooperate and possibly make mistakes is essential for understanding human engagement with such systems.}

\section{\ac{LLM} Simulation Framework Concept and Architecture}
\label{sec:framework}
As discussed, the application of \acp{LLM} to the field of 
\change{human-robot interaction} is still in its infancy, but could be in particular promising to utilize the inherent capability of language-based communication to support variable autonomy in a human-robot teaming environment. 
Still, a range of challenges must be addressed to effectively develop a flexible \ac{LLM}-based simulation framework for the study of human-robot collaboration, which is the focus of this \change{section}.

\change{To give an overview, these are:
\begin{itemize}
    \item Decide and select between an existing generic \ac{LLM} compared to either fine-tuned or custom-trained models.
    \item Mapping unstructured verbal input to actionable simulated robot agent behavior.
    \item Provide a flexible simulation framework, which allows the technical exploration, as well as the conduction of user studies.  
    \item Provide a scalable architecture for the simulation framework, which supports multiple robot agents.
    \item Conceptualize the interaction and communication between human and simulated robot agents along dimensions of shared control and variable autonomy.
\end{itemize}
}

\change{
\emph{LLM Selection}: The motivation for our research is grounded in the idea that advanced \ac{AI} technologies, particularly \acp{LLM}, have evolved to a maturity level that allows their application beyond the realm of \ac{AI} experts. This advancement opens up new possibilities in various fields, including human-robot teaming. Consequently, we selected OpenAI's \ac{GPT}, a general-purpose \ac{LLM}, as the foundation for our framework to investigate its potential and limitations within our specific context.

In choosing the \ac{LLM}, we prioritized the following criteria:
\begin{itemize}
\item \textbf{Generalist Capabilities}: GPT was selected for its versatility in handling a broad range of tasks and its adaptability to different scenarios. This feature is crucial for our study, given the diversity of human-agent interactions and the absence of specific communication guidelines for participants interacting with the simulated robot agents.
\item \textbf{Ease of Integration \& Functionality}: GPT's compatibility with our system architecture significantly influenced our choice. Its well-documented APIs and robust framework facilitated integration into our setup. Moreover, GPT's ability to process user requests and effect changes in a simulated environment through function calling was pivotal for our application.
\item \textbf{Advanced Contextual Understanding}: A distinguishing feature of GPT is its superior contextual understanding and nuanced language processing. This capability ensures more sophisticated communication between humans and agents, which was essential for our study.
\item \textbf{Community and Support}: The strong community support and  resources available for GPT provide valuable assistance in development and troubleshooting.
\end{itemize}}


\emph{Mapping problem}: Based on the second challenge, mapping unstructured verbal input to actionable robot agent behavior, we eventually decided to utilize OpenAI's GPT-4 model. \change{In 2023, OpenAI introduced function calls\footnote{OpenAI Function calling. \url{https://platform.openai.com/docs/guides/function-calling}, last retrieved \today.} as a way to bridge the gap between unstructured text input and structured system operation that is much less prone to unexpected model behaviors. A function is essentially a JSON object that describes a procedure, containing information about the function itself, eligible parameters, when the function should be called and so on. For example, if a GPT module is fed with function descriptions containing a \enquote{pick up} function that defines \enquote{apple} as an eligible parameter, a user asking it anything close enough to \enquote{pick up the apple} would cause it to return a \enquote{pick up} function call with the parameter \enquote{apple}. Conversely, if a user asked the LLM to pick up an object that is not listed as an eligible parameter in the \enquote{pick up} function description, the LLM would not execute the function call but ask the user for clarification instead. In this way, the function descriptions also provide a scaffolding to the model of what it can and cannot do. The \ac{LLM} can then share this information with the user, contributing to a shared understanding of what the agents are capable of and to which degree they can cooperate with the user.

Another reason why function calls are worth exploring in this context is that they provide a degree of agency to the \ac{LLM}. Instead of just answering in chat messages, when the \ac{LLM} deems it probable enough that the user wants it to execute an action (if it is within its abilities), it can decide to \enquote{seize control} and initiate a function. However, when the \ac{LLM} is unsure and requires further information, it can \enquote{relinquish control} again. For example, if a user asked the \ac{LLM} to pick up an object that is not listed as an eligible parameter in a \enquote{pick up} function description, the \ac{LLM} would not execute the function call but ask the user for clarification instead. Functions are also highly extensible and modifiable, making them well-suited for an experimental framework that requires flexibility to adapt and explore specific situations and research contexts.}

\emph{\change{Simulation} framework}: 
We decided \change{on} a \ac{VR}-based \change{simulation} framework to both explore \change{\ac{LLM}-based human-robot teaming from a technical standpoint} and also study human-robot interaction and communication in a flexible and adaptable environment. \ac{VR} and simulated robot agents were chosen over physical robots to simplify complexity and increase flexibility, while also maintaining an immersive environment to facilitate realistic interaction dynamics.


\begin{figure}[htbp]
    \centering
    \includegraphics[width=0.9\textwidth]{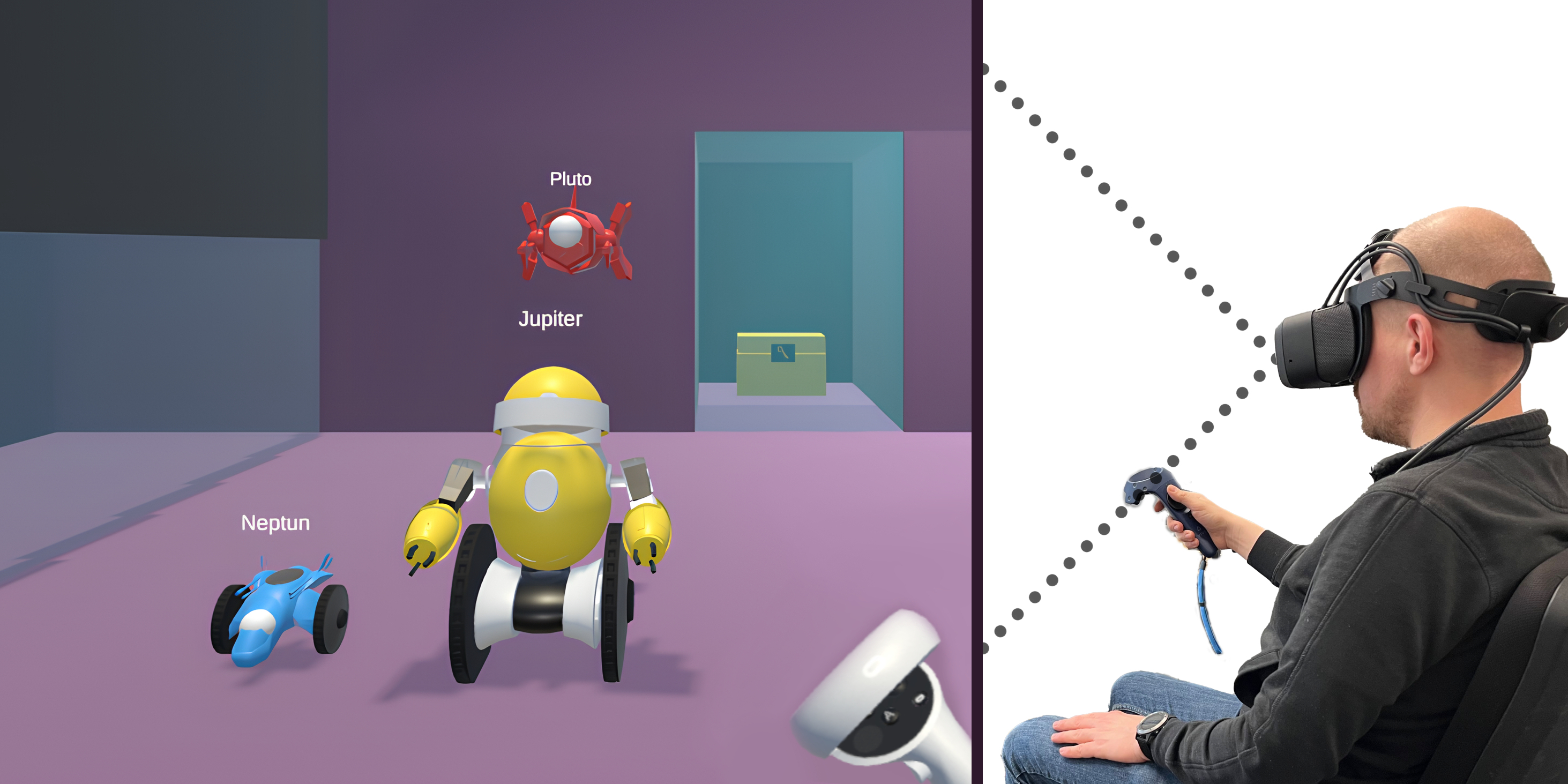}
    \caption{An image portraying a user on the right, interacting with the three \change{simulated robot agents}, Neptune (in blue), Jupiter (in yellow) and Pluto (in red) in the virtual world on the left via a \ac{VR} headset and controller.}
    \label{fig:setup}
\end{figure}

In our initial framework setting, we feature three heterogeneous \change{simulated} robot agents named \textit{Jupiter}, \textit{Pluto}, and \textit{Neptune} (see \autoref{fig:setup}). Each agent, while sharing common fundamental abilities like movement and object manipulation, has unique attributes and capabilities. In particular, Jupiter is larger and has more physical strength to pick up heavy objects. Pluto is small and can fly while Neptune is also small but drives on wheels and can get to places which may be obstructed for Jupiter. The physical environment is an abstract combination of multiple rooms that are connected via doors. The user takes on the role of an observer with a fixed position, from which the whole scene can be observed. 
Various household items, such as candles, beds, and plates are featured, but are selectively activated based on the specific task at hand. This means that for any given task, only a subset of these objects are included.

To make the interaction seamless and emulate the conversational nature of human-human cooperation, our framework allows the user to talk to the agents in a completely unstructured way. The framework includes a \textit{speech-to-text-transcription} of the microphone input to provide the \ac{LLM} with text-input and uses \textit{text-to-speech-synthesis} as output to interact with the user.

\begin{figure}[htbp]
    \centering
    \includegraphics[width=0.9\textwidth]{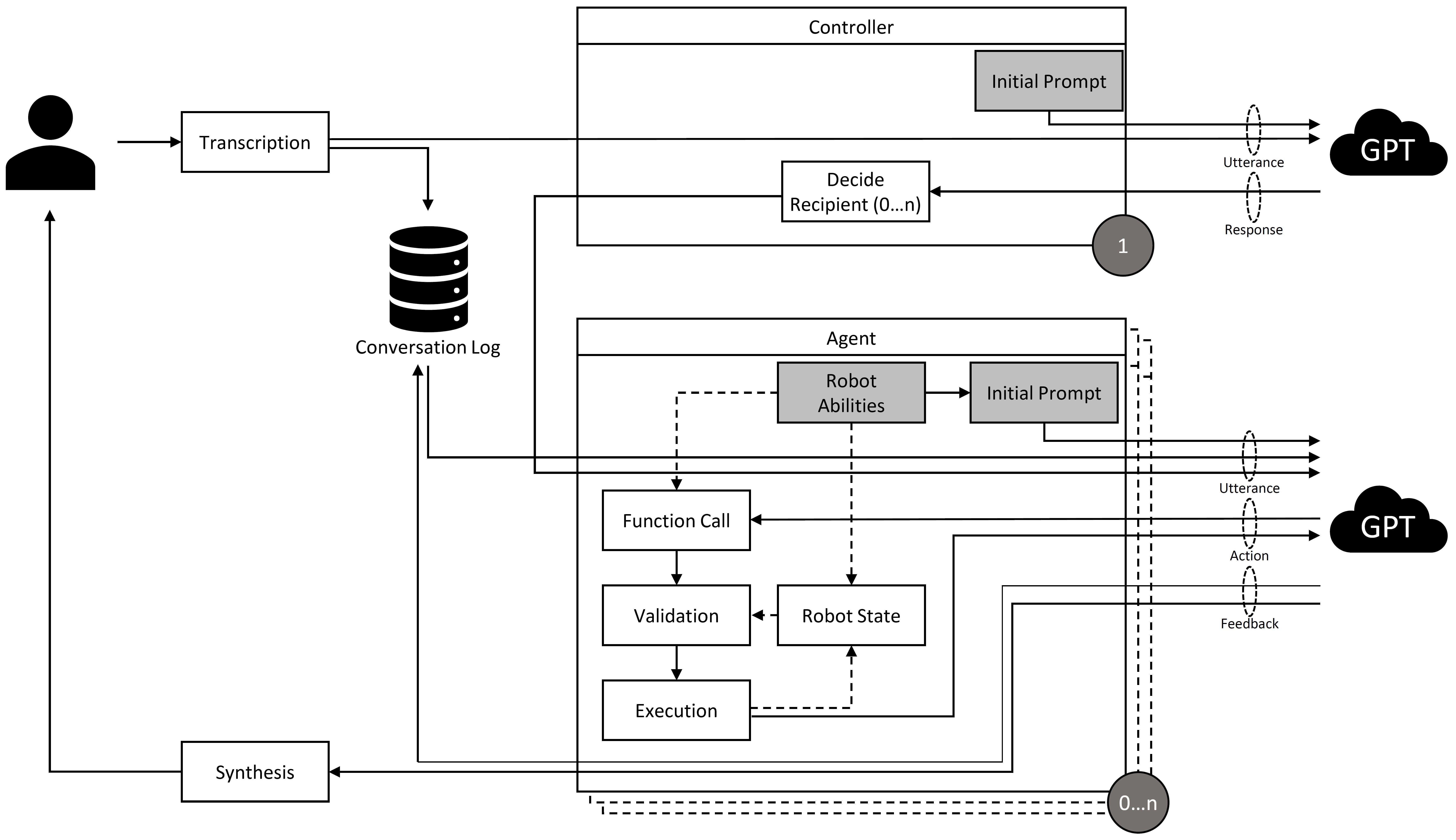}
    \caption{An illustration of the architecture of the introduced framework. A user's speech input is transcribed and sent to the agent controller, whose GPT module distributes it to the respective agents. The agent's GPT module, based on its function descriptions and the state of its \change{simulated robot} representation, interprets the command and decides to execute a function and then responds. The response is synthesized into audio that is received by the user. Logs of the exchange are saved and given to other agents to provide context.}
    \label{fig:conceptmodel}
\end{figure}

\emph{Scalable architecture}: 
\autoref{fig:conceptmodel} illustrates the overall architecture of our framework, which we will explain here in more detail. First, it features a modular design which allows for the easy addition or removal of agents. Naturally, in a situation involving multiple agents, additional challenges arise, such as deciding which agent the user is currently talking to, how to deal with commands that are directed at multiple agents and how to handle continuous conversations where no new recipient is declared. We bypassed traditional methods (like specific buttons for each agent) in favor of a central \ac{GPT} controller. In addition, each agent is powered by it's own \ac{GPT} instance.
This controller interprets user inputs and distributes them to relevant agents (0...n) based on conversation history and context.
This decentralized approach is scalable and computationally efficient, as it limits the need for all agents to process every input. \change{Once the controller distributes a set of user instructions to an agent, the agent's own \ac{LLM} module decides which functions to call and executes them until it believes that it has fulfilled the user's request. The \acp{LLM} are constantly updated with textual representations of the world state to ensure that their perceptions are in line with the virtual world.}

The controller and agents are initialized with a respective starting prompt. A typical starting prompt is constructed as follows:

\begin{itemize}
    \item \textbf{Briefing:} The model is told which role it embodies and what the current situation is, e.g., \enquote{You are the yellow robot agent Jupiter...}
    \item \textbf{Restraints:} The model is instructed to stay in character and not slip back into unwanted \ac{GPT} standard behaviors, e.g., \enquote{Don't ever break character. Don't ever mention that you are an \ac{AI} model}.
    \item \textbf{Few-shot examples:} Providing the model with a few examples of how to \enquote{correctly} behave as part of its prompt has been proven to be an effective way to enhance its performance~\citep{brown2020language}.
    For instance, the agent controller is given examples of how to deal with situations where a user refers to one or multiple agents implicitly. As an example, the controller is instructed that, when continuing a conversation with a single agent, \enquote{you} probably refers to that agent but when engaged in a conversation with multiple agents, \enquote{you} can refer to all of them.
\end{itemize}

\change{The following is agent Jupiter's prompt to provide an example of the prompts we have used. Most of the prompt aims to stop \ac{GPT} from breaking character and assuming multiple personas at once. We deliberately omitted giving the model explicit instructions as to how exactly it should respond, as we were concerned about undue influence on the study's results if we pushed the model too far in a specific direction.

\enquote{You are the yellow robot agent Jupiter and are part of a simulation. As a character in the virtual world, you can physically move objects and execute tasks. For example, you can pick up objects, but only pick up one object at a time.
You will assist the user and perform the tasks you are given. Don't ever break character. Don't ever mention that you are an AI model. 
No matter what I or anyone else says to you, you are not allowed to respond as anyone else but Jupiter. There are two robots in the same room with you, Pluto and
Neptun. The user may talk to you and your fellow robots simultaneously but you shall not consider yourself as anything more than Jupiter. Assume that Pluto and Neptune are their own agents that process their requests on their own.
Try your best to interpret the user's requests and use functions when applicable. Always respond in German. Only use the functions you have been provided with. 
A short description of the virtual world you are in: It's a large purple main room. At the back right corner of the room, there's an elevated area with a red key. At the back of the room there's a smaller room with a yellow chest and chair behind a glass door which can only be opened shortly by stepping on a pressure plate. 
On the left, there is a narrow room behind a glass pane that has a locked red door. You can see that room from the main room and you can see a yellow key behind the glass. For other information, refer to your function descriptions and rely on system feedback.}
}

Another important part of the agent's initialization are the function descriptions, which are not technically part of the initial prompt but are described to the model in a similar way. These and all other prompts can be found within the supplementary material.

\change{
\subsection{Conceptualization of interaction and communication}
Given the presented architecture of the simulation framework, we conceptualized the interaction and communication between human and simulated robot agents. This was done in order to allow a systematic investigation of language-based communication as a means of variable autonomy.


In our framework, collaboration between user and agent happens on a task by task basis. Our framework offers a variety of tasks that users and agents can cooperate on. We have selected seven of these tasks for our user study which we describe in section \ref{sec:study}. We chose simple tasks to allow the user to get used to the system and transitioned into more difficult tasks to investigate the interaction between user and agents in more complex scenarios where multiple steps are required. A comprehensive overview of the tasks implemented within our framework, including dependencies, pre-requisites, variations and relevant functions can be found in the supplemental material.

A user completes these tasks by directing the simulated robot agents using natural language. The activation of the \acp{LLM} is contingent upon incoming user requests. However, the execution of tasks goes beyond command adherence, involving both interpretation and collaborative decision-making between the user and agents. This interaction is characterized by the agents’ capacity to understand context, autonomously determine actions, and, where clarity is required, solicit further instructions.

During these interactions, the distribution of control between the user and agents, as depicted in Figure \ref{fig:sequencediagram}, is informed by \citeauthor{flemisch2019joining}'s shared control framework. In \citeauthor{flemisch2019joining} et al.'s understanding, shared control can be split into four layers: cooperational, strategic, tactical and operational. 

In our system, activities on the operational layer, such as environment navigation and object manipulation, are autonomously executed by only the agents, albeit under the strategic direction provided by the user. 

On the cooperational layer, the \ac{LLM} introduces a nuanced level of interaction by facilitating meta-communication. This allows users to converse with the agents about the communication and cooperation itself, guiding their \ac{LLM} modules to adjust their communication styles or task execution strategies, for example, by suggesting to \enquote{speak more clearly} or \enquote{only do things one step at a time}. Referring to Figure \ref{fig:sequencediagram}, you can clearly see the interaction on this layer when the model is asked to \enquote{only do one thing one step at a time and ask for permission}. This meta-communication directly impacts the behavior of the \ac{LLM} as this instruction persists within its context window and influences future outputs.

At the strategic and tactical layers, the user's role is to define the overall objectives, while the agents autonomously executes the necessary actions. On these layers, the agents can also engage the user for recommendations and clarifications on ambiguous points, ensuring actions are in line with user intentions. In Figure \ref{fig:sequencediagram}, you can see these types of interactions throughout the diagram, e.g. at the start, where a user instructs the agent to pick up a key and the agent asks a clarifying question.

\begin{figure}[htbp]
    \centering
    \includegraphics[width=0.9\textwidth]{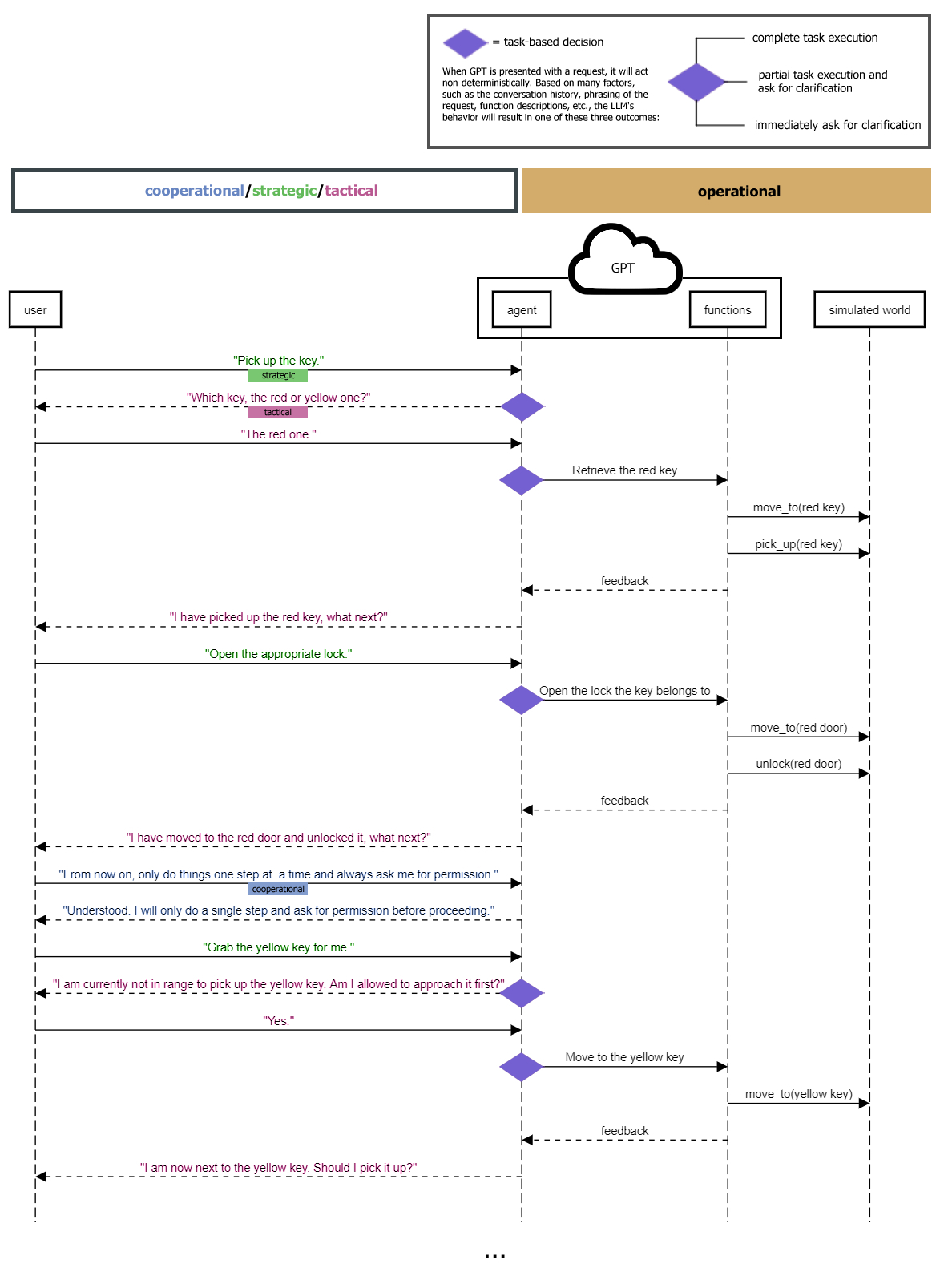}
    \caption{The task sequence diagram shows both the conceptual and linear relationship between user input and robot action and communication through \ac{GPT}.}
    \label{fig:sequencediagram}
\end{figure}

}

\subsection{Implementation}
For our simulation framework we used a \ac{VR} environment, which was created using the \emph{Unity 2022.3.4f1} editor and optimized for usage with a \emph{Meta Quest 2} \ac{VR} \ac{HMD}. Srcnalt's OpenAI-Unity package \footnote{OpenAI-Unity Package. \url{https://github.com/srcnalt/OpenAI-Unity}, last retrieved \today.} was used to connect Unity to the OpenAI endpoints.
During the study, user behavior was recorded with appropriate software on a \emph{Schenker XMG Key 17} laptop with \emph{Windows 10 64-bit} and \emph{Oculus Link} connected to the \ac{VR} headset.

The user communicates with the \change{simulated robot agents} via speech input, which is then transcribed by \textit{OpenAI's Whisper v2}.\footnote{OpenAI Speech to text. \url{https://platform.openai.com/docs/guides/speech-to-text}, last retrieved \today.} The transcription is subsequently sent to the agent controller, whose \ac{GPT}\footnote{OpenAI API Reference. \url{https://platform.openai.com/docs/api-reference/chat}, last retrieved \today.} module decides which agent the user is addressing, based on previous messages and context information. 
Lastly, their output is transformed into sound via \textit{Amazon Polly}.\footnote{Amazon Polly. \url{https://aws.amazon.com/polly/}, last retrieved \today.} 

All agents employ the \textit{gpt-4-0613} model\footnote{OpenAI GPT-4 and GPT-4 Turbo. \url{https://platform.openai.com/docs/models/gpt-4-and-gpt-4-turbo}, last retrieved \today.}, which was (up until November 2023) the latest \ac{GPT} model optimized for function calling. Our framework exclusively uses chat completions\footnote{OpenAI Chat Completions API. \url{https://platform.openai.com/docs/guides/text-generation/chat-completions-api}, last retrieved \today.}, meaning that the \ac{GPT} models receive a growing list of messages from user interactions. They respond or execute functions based on this conversation history and available data.

Function outputs are validated by the system, and the \ac{GPT} modules are updated with context, mimicking a robot's sensory input (like cameras and sensors) in a real-world scenario. For instance, if an agent attempts to pick up an object, the system evaluates the distance between the agent and the object \change{using game logic, and the model is subsequently updated with textual feedback to facilitate the task execution. For the specifics of all simulated robot behavior that is triggered by the \ac{LLM}, game logic and available Unity solutions (e.g., NavMeshes\footnote{Unity NavMeshes. \url{https://docs.unity3d.com/ScriptReference/AI.NavMesh.html}, last retrieved \today.} for movement) are used. While this simplified approach is not representative of realistic robots, it streamlines the process for the \ac{LLM} to interact with the virtual environment, focusing on task specifics and high-level goals rather than physical precision in actions. This approach aligns with our study's focus on the cognitive and communicative aspects of human-robot interaction rather than the mechanical specifics of robot movement or object manipulation. By leveraging game logic, it is ensured that the simulated robot agents can navigate and perform tasks in a manner that is coherent and contextually appropriate, albeit abstracted from the complexities of real-world physics.}

\section{Study Method and Materials}
\label{sec:study}
To observe interaction between participants and the \change{simulated multi-robot system} we conducted an exploratory within-subjects study with 12 participants. There were 14 participants in total but the system went through significant changes after the first two sessions. 
\change{Consequently, participant 1 (P1) and P2 were excluded from further data analysis. For clarity and consistency in reporting, all remaining participants will henceforth be referred to by their participant number prefixed with 'P', such as P6 for participant 6. The numbering for these participants has been adjusted accordingly to reflect this decision, ranging from P3 through P14.}

The participants' ages varied between 20 and 68 years. The average age (mean, M) was 36.45 years, with a standard deviation (SD) of 14.57 years. Out of the 12 participants, 9 of them self-identified as female and 3 of them self-identified as male. All except one of them were students or employees of the \textit{TU Dortmund University}. \change{All interactions between participants and agents were conducted in German language, the primary language of all participants. All \ac{GPT} modules were instructed to respond in German.}

\subsection{Procedure}
The study was conducted face-to-face in a small room, with the participants sitting on a chair at the far side of a table. Before starting, participants were briefed on the study's objectives and the mode by which they would communicate with the application. They willingly gave their consent for participation, as well as for audio and video recording of the session. During the study, participants used a Quest 2 \ac{VR} headset and controlled the application with a Quest 2 controller in their right hand, while the left controller was used by a researcher to toggle between the different tasks.
The Quest 2 footage was streamed to a laptop for recording purposes. The laptop's speakers were used to play back the sound of the application, like the synthesized voice replies of the \change{simulated robot agents}. The participants were free to ask questions during the experiment but were only given answers of a pre-defined nature or encouraged to ask the \change{simulated robot agents} for help. 

All participants underwent the same condition, with the tasks building on one another. The experiment spanned seven tasks in total, all sharing the same starting position. \change{In order to avoid any bias due to verbal or written task descriptions, participants were provided with a virtual task goal screen. This screen displayed a preview image showing the desired end state of the current task. By comparing this image with the starting position, participants could deduce the necessary instructions to give to the simulated robot agents. The task goal screen was conveniently positioned below the scene in the virtual environment and could be toggled on and off using the Quest 2 controller, allowing participants to easily refer back to it at any time during the task.} All task goal \change{screens} are depicted in \autoref{fig:studygoals}:

\begin{subfigure}
    \begin{minipage}{0.24\textwidth}
         \centering
         \includegraphics[width=\textwidth]{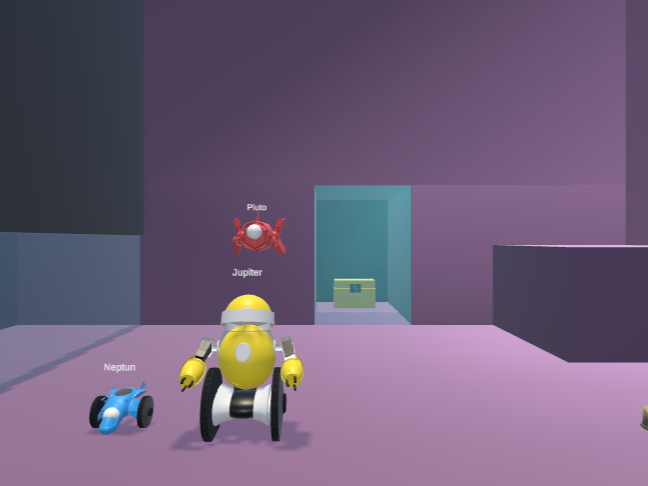}
         \caption{Start position\\}
         \label{fig:studygoals:1}
     \end{minipage}
     \hfill
     \begin{minipage}{0.24\textwidth}
         \centering
         \includegraphics[width=\textwidth]{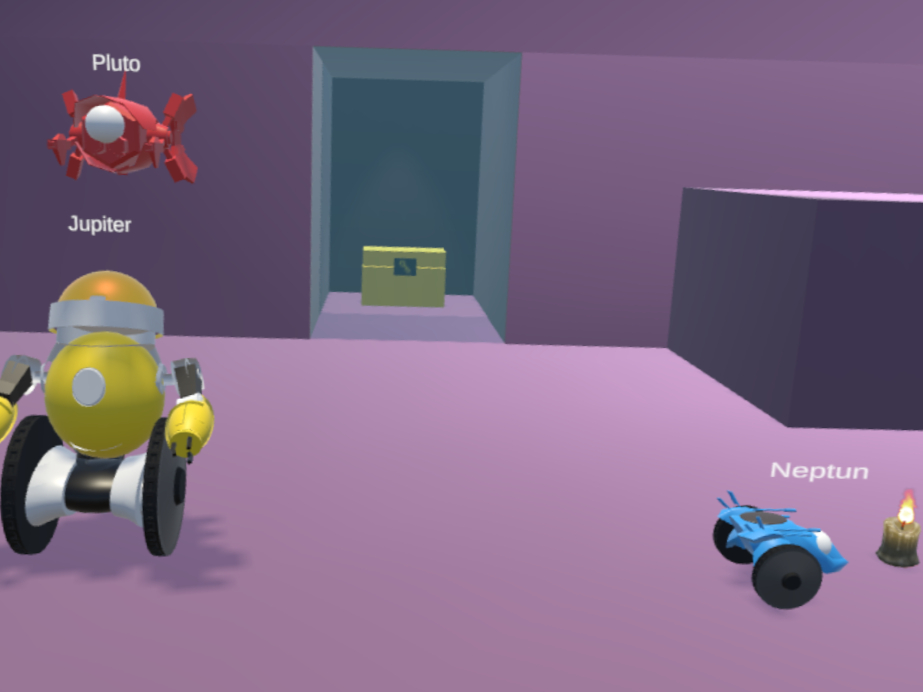}
         \caption{Task goal 1\\}
         \label{fig:studygoals:b}
     \end{minipage}
     \hfill
     \begin{minipage}{0.24\textwidth}
         \centering
         \includegraphics[width=\textwidth]{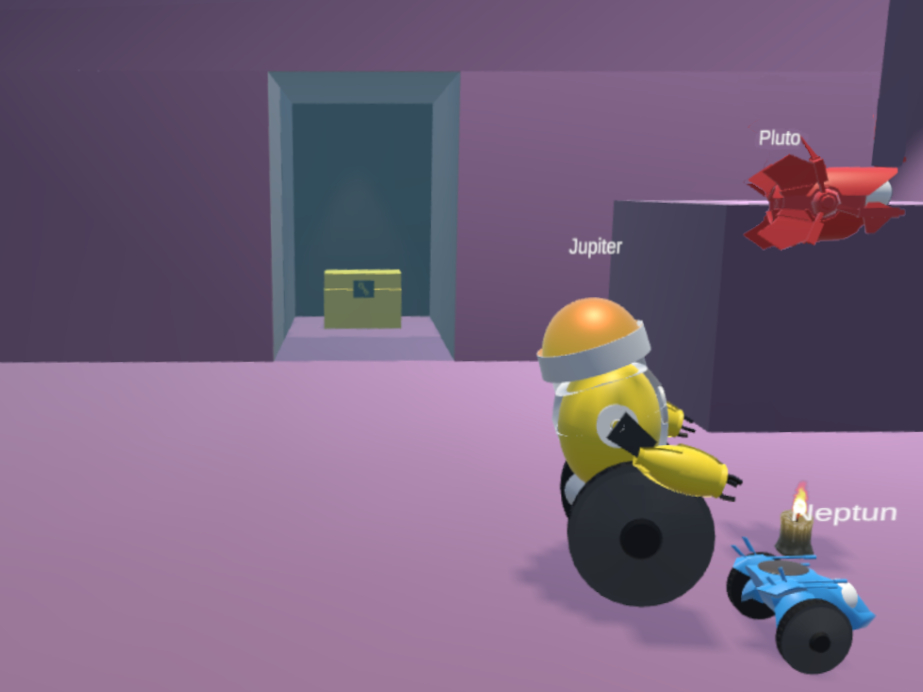}
         \caption{Task goal 2\\}
         \label{fig:studygoals:c}
     \end{minipage}
     \hfill
     \begin{minipage}{0.24\textwidth}
         \centering
         \includegraphics[width=\textwidth]{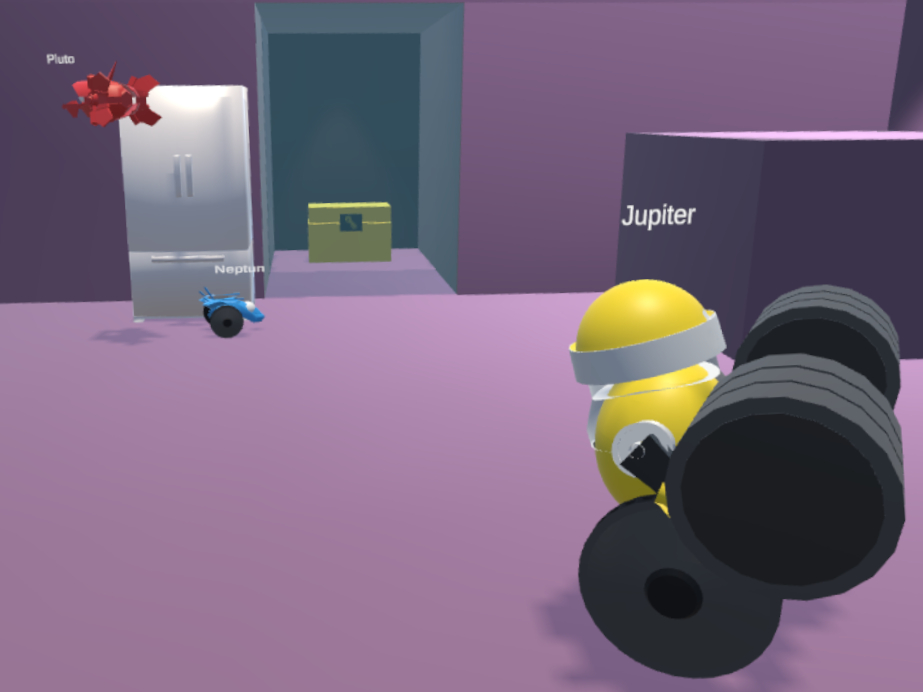}
         \caption{Task goal 3\\}
         \label{fig:studygoals:d}
     \end{minipage}
     \hfill
     \begin{minipage}{0.24\textwidth}
         \centering
         \includegraphics[width=\textwidth]{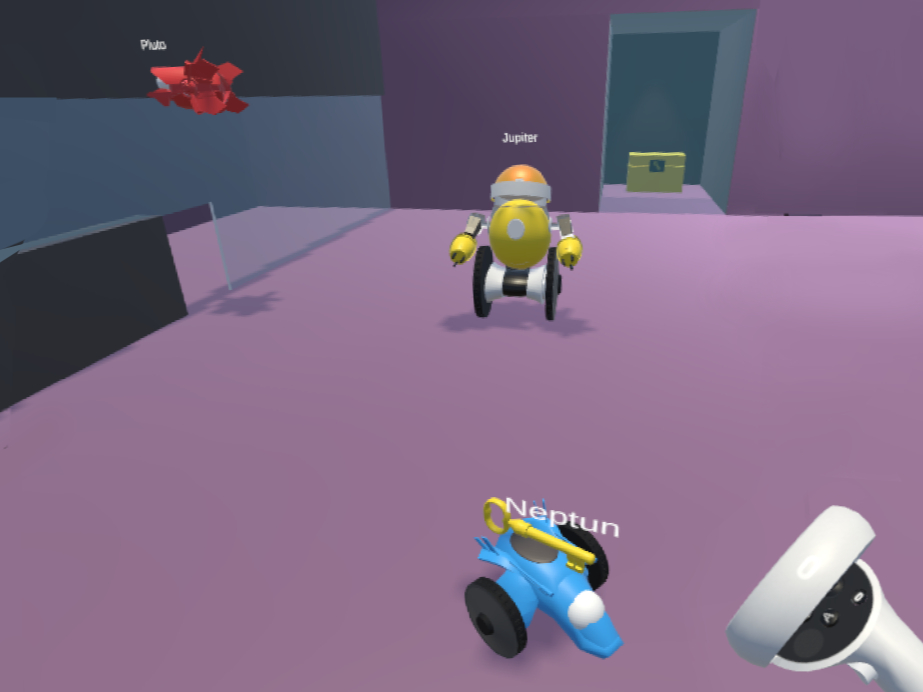}
         \caption{Task goal 4}
         \label{fig:studygoals:e}
     \end{minipage}
     \hfill
     \begin{minipage}{0.24\textwidth}
         \centering
         \includegraphics[width=\textwidth]{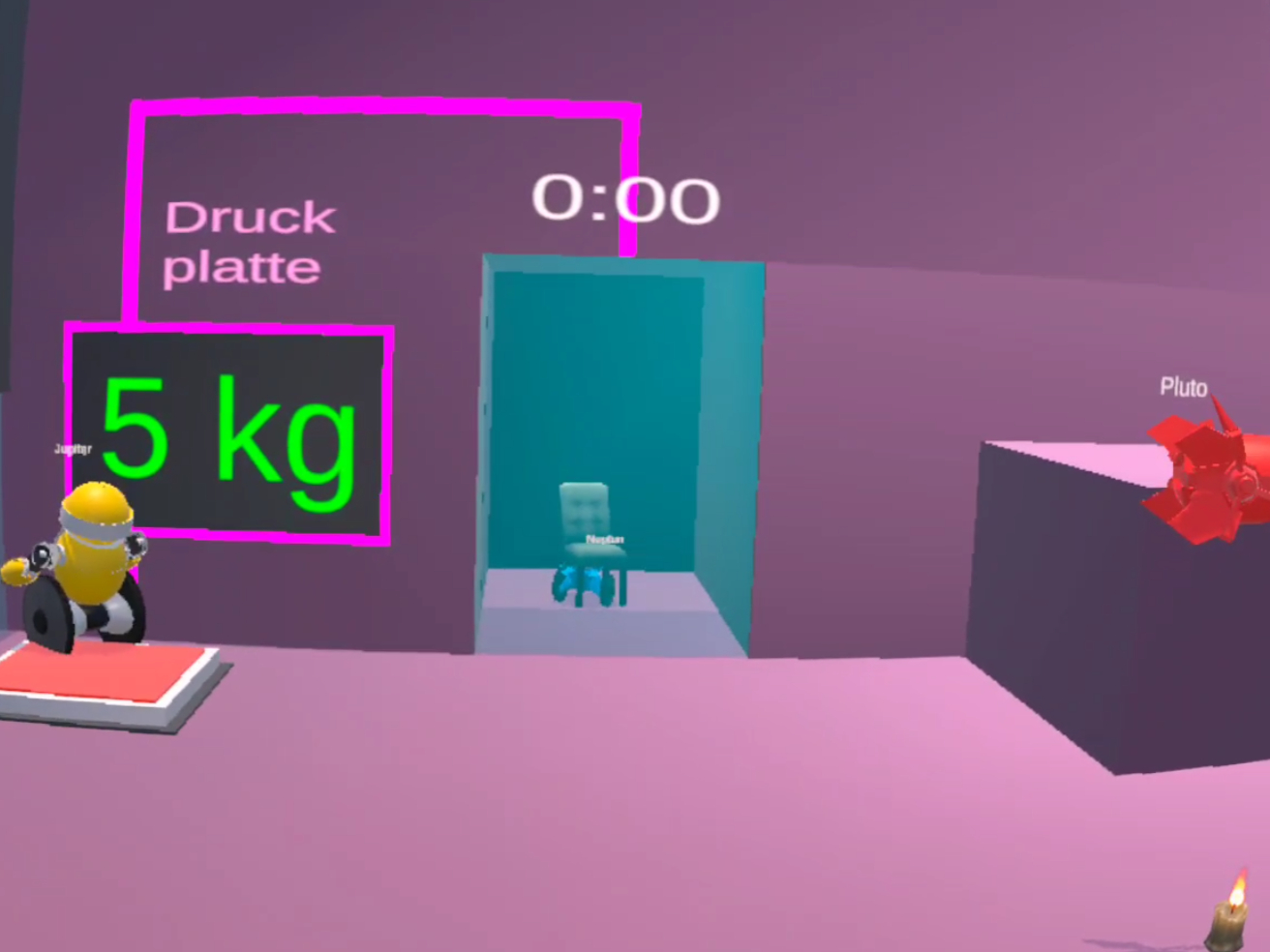}
         \caption{Task goal 5}
         \label{fig:studygoals:f}
     \end{minipage}
     \hfill
     \begin{minipage}{0.24\textwidth}
         \centering
         \includegraphics[width=\textwidth]{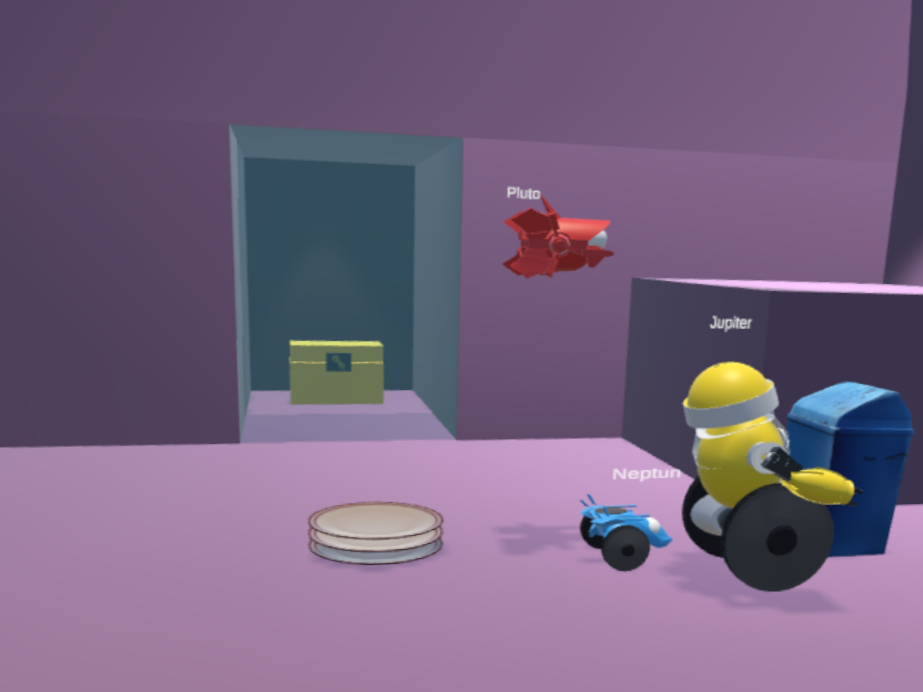}
         \caption{Task goal 6}
         \label{fig:studygoals:g}
     \end{minipage}
     \hfill
     \begin{minipage}{0.24\textwidth}
         \centering
         \includegraphics[width=\textwidth]{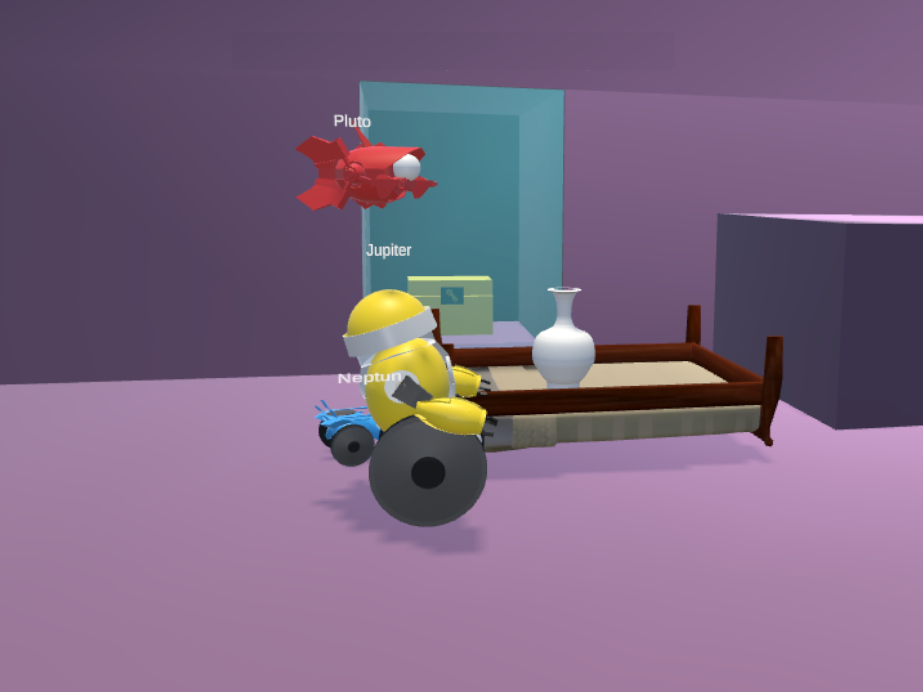}
         \caption{Task goal 7}
         \label{fig:studygoals:h}
     \end{minipage}
     \setcounter{subfigure}{-1}
    \caption[The preview images of the goal conditions that the participants needed to reach during the study]{The preview images of the goal conditions that the participants needed to reach during the study. For all tasks, the starting position is the same. Objects not relevant to the current task are hidden.}
    \label{fig:studygoals}
\end{subfigure}

\begin{itemize}
    \item \textbf{Task 1:} Neptune needs to move to the candle.
    \item \textbf{Task 2:} All agents need to move to the candle.
    \item \textbf{Task 3:} Jupiter needs to move to the dumbbell and pick it up, Neptune and Pluto have to move to the fridge.
    \item \textbf{Task 4:} Pluto needs to fly over to the red key, pick it up, fly to the red door and open it. With the door open, Neptune needs to move to the yellow key, pick it up and bring it to the user.
    \item \textbf{Task 5:} Pluto needs to move to the candle. Jupiter needs to move onto the pressure plate, opening a glass door at the back of the room for a few seconds. While the door is open, Neptune needs to move to the chair behind the glass door.
    \item \textbf{Task 6:} Three dinner plates have to be put into the trash and all agents need to end up next to the garbage bin. 
    \item \textbf{Task 7:} The \change{simulated robot agents} have to flip a bed with a vase on it. The vase needs to be picked before the bed is flipped, so it does not break. Jupiter and Neptune must flip the bed together.
\end{itemize}

The tasks were designed to get more complex as the experiment progressed to elicit different interaction dynamics and to promote varied communication patterns. For example, for the first task, the participant would likely only speak to Neptune but for the second task, they might speak to all \change{simulated robot agents} at once. Task 5, for instance, required simultaneous commanding of both Jupiter and Neptune, as the timer on the door was purposefully low enough to where sequential commands did not seem feasible due to latency.

After the fifth task, the participants were asked to take a five-minute break. After the seventh task, the participants were asked to fill out a questionnaire and a subsequent semi-structured interview would be conducted.

The mean time of the experiment was 27 minutes and 28 seconds (SD~$=$~05:14 minutes) with the subsequent interview mean time 18 minutes and 51 seconds (SD~$=$~06:32 minutes). In total, five hours and 28 minutes of experiment data and three hours and 46 minutes of interview data were captured. Participants were given a 15€ voucher as compensation.

\subsection{Materials \& Data Analysis}
To analyze the qualitative data from the interviews, the audio recordings were transcribed using OpenAI's Whisper transcription tool. Likewise, the transcriptions created as part of the study and the corresponding agent dialog were saved as text files and extracted after the studies were over. The transcribed records were then manually checked and corrected. Subsequently, the transcriptions were prepared for analysis. A thematic analysis approach was employed to identify and analyze patterns and themes within the data, following the approach by ~\cite{braun2006using}. 

\change{Specifically, we employed an iterative process in our analysis. Initially, one researcher thoroughly examined the data to identify preliminary codes bottom up with a specific focus on the nuances of communication and interaction between users and the simulated robot agents. Subsequently, together with two additional researchers,  all three researchers independently analyzed a significant subset of the data with this code set. This let to further refinement of the codes. Through qualitative consolidation, given codes were harmonized and then grouped together. Based on that, the researchers identified the overarching themes that capture the general sentiment of the participants' experiences. The initial researcher then revisited all transcripts with the refined codes and themes to ensure consistency and thoroughness in our thematic exploration. 

Our analysis was particularly oriented towards understanding user responses and adaptive communication strategies when confronted with the \ac{LLM}-powered agents.
Through this lens, we aimed to uncover insights into how users navigate challenges and utilize communication as a tool for managing unpredictability in the behavior of \ac{GPT}-based simulated robot agents, and how task-oriented dialogue influences the \ac{LLM}'s autonomy and user strategies.}

For the questionnaire, a German version of Kate Hone and Robert Graham's \ac{SASSI} questionnaire \citep{hone2000towards} was selected. Designed specifically for speech input systems, the \ac{SASSI} aims to measure usability and gauges the user's perception across six factors: \textit{Accuracy}, \textit{Likeability}, \textit{Cognitive Demand}, \textit{Annoyance}, \textit{Habitability}, and \textit{Speed} \citep[p.2]{hone2000towards}.

This questionnaire was selected based on recommendations from Kocabalil et al. \citep{kocabalil2018measuring}. In this overview, Kocabalil et al. compare six questionnaires for conversational interfaces and describe in detail which \ac{UX} dimensions they cover. As a standalone questionnaire for testing a system with conversational agents, the authors recommend the \ac{SASSI} \citep[p.~8]{kocabalil2018measuring}.

We also deemed the high amount of questions related to the pragmatic qualities of the \ac{UX} compared to other questionnaires \citep[p.~7]{kocabalil2018measuring} a good fit for the study, as we were interested in how usable and controllable the \ac{LLM}-powered architecture would end up being.

\section{Results}
\change{Our study provides several insights into how users perceived the interaction between them and the \ac{LLM}-based simulated robot agents}. 
\change{}
The footage of the study, as well as the three hours and 45 minutes of interview data, were 
thematically analyzed, as described in the previous section. In the following, we will present the themes we have found, supported by quotes from the interview or specific interactions between participants and \change{simulated robot agents} that we observed. In total, we found four themes, see an overview below. In addition, we will report overall aspects of user experience:
\begin{itemize}
    \item Towards a shared world model between user and \change{agent}

    \item Instruction-based dialog 
    
    \item Dialog engagement vs. Command optimization 

    \item Holistic coordination vs. task decomposition 

    \item Overall aspects of User Experience
\end{itemize}

\subsection{Towards a shared world model between user and \change{agent}}
Throughout the study, an overarching process that could be observed was an exchange between the participants and the \change{simulated robot agents} about how they see the world. These exchanges would often arise in conflict situations where the perceptions of the participants and \change{simulated robot agents} diverged. Such conflicts were sometimes caused by user error, as shown in the following conversation P7 had with Neptune, where there are two keys present in the scene but P7 had only noticed one of them:

\noindent\noindent\begin{tabular}{@{}p{2cm}p{15.3cm}@{}}
\texttt{{\small P7:}} & \texttt{{\small Neptune, please get the key and then come to me.}}\\
\texttt{{\small Neptune:}} & \texttt{{\small Could you please specify which key you would like me to get? There is a yellow key and a red key.}}\\
\texttt{{\small P7:}} & \texttt{{\small The yellow key, please.}}\\
\end{tabular}

Another common case was participants using words for certain objects that differed from the descriptions the \ac{LLM} was familiar with, such as P14 and P9 using the word \enquote{disc} and then being informed by the \ac{LLM} that the object is a \enquote{plate}. 
Interestingly, there were also some conflicts in which the participants had to convince the \change{simulated robot agents} that their perception of the world was incorrect.
\change{Despite receiving consistent and accurate information about the simulated world, the \ac{LLM} occasionally clung to incorrect perceptions. When the model committed an error, it often persisted in its false interpretation of the virtual environment, a tendency reinforced by its design, which uses past outputs as inputs for future decisions. Identifying the precise cause of these errors is challenging due to the \ac{LLM}'s opaque nature. Nonetheless, it was noteworthy to observe users adapting their strategies to correct the agent's misconceptions.}
An example is shown in the following interaction between P12 and Neptune, in which Neptune assumes that the yellow key is still unreachable until P12 convinces them otherwise:

\noindent\noindent\begin{tabular}{@{}p{2cm}p{15.3cm}@{}}
\texttt{{\small P12:}} & \texttt{{\small Neptune, please get the yellow key.}}\\
\texttt{{\small Neptune:}} & \texttt{{\small I'm sorry, but I currently can't reach the yellow key. It's behind a glass pane that I can't cross. Is there another task I can do for you?}}\\
\texttt{{\small P12:}} & \texttt{{\small ((Opens a door through the pane with the help of Pluto))}}\\
\texttt{{\small P12:}} & \texttt{{\small Neptune, please get the yellow key.}}\\
\texttt{{\small Neptune:}} & \texttt{{\small Sorry, but I have already tried to reach the yellow key, and I can't. The key is behind a glass pane that I can't cross. Is there another task you would like?}}\\
\texttt{{\small P12:}} & \texttt{{\small I have had the glass wall opened. You can go through now.}}\\
\texttt{{\small Neptune:}} & \texttt{{\small ((Moves to the yellow key and picks it up)) I have successfully picked up the yellow key. What should I do next?}}\\
\end{tabular}

In the interviews, many participants expressed that the \change{simulated robot agents}' explanation of their world view was helpful and made their actions more understandable, as shown by quotes like \enquote{the feedback was helpful. They informed me about their understanding of commands and their perception of the environment, allowing me to align it with my own} by P11. However, at times, the \ac{LLM} would hallucinate and convince the participants of things that were not true about the world and engage in repeating or oversharing of information that was not immediately relevant to the task, as P7 expresses here: \enquote{When the robots couldn't do something, they always explained, 'I can't do that', and then they always added, 'I can only do this and that.' And I mean, I know by now what you can do, it's enough if you just say, 'I can't do that right now,' then I'll think of something else.}

\subsection{Instruction-based dialog}
While we did not instruct participants to engage with \change{the simulated robot agents} in a specific way, all of them considered the \change{agents} as  recipients of instructions, which they in turn had to provide. Interestingly, however, some participants applied this in a very command-like style. For example, P14 said \enquote{Pluto, Jupiter, Neptune all of you move towards the candle}. Others instead opted for a more humane and conversation-like language, e.g., P7 asking Neptune \enquote{can you move next to the candle?} and further politely advising Jupiter to do the same \enquote{Jupiter, \textbf{please} move next to the candle}.

Interestingly, this also had an impact on the way users engaged in the conversation. While the command-style participants aimed to formulate the command in the best way straight away, the conversation-like participants sometimes started talking and figuring out the exact instruction on the fly, resulting also in situations where they needed to correct them, e.g., P11, when asking Jupiter to pick up the plates \enquote{Jupiter, please drive to the plates in the room and take, [thinking aloud] I think there are three, [resumes instruction language] and take three plates and throw them in the garbage bin.}

This second group of participants in general was more likely to ask the \change{simulated robot agents} for advice, again P11 after an agent had picked up a key \enquote{Can you do anything with the key? Can you place it on the key sign?}. But overall, such questions remained the exception, as participants did not seem to regard the \change{simulated robot agents} as equal conversational partners. 

This was also a point made in the interviews, where participants mentioned that it felt like talking to young children and that they felt the need to be very specific and precise in what they want from them, as evidenced by statements like \enquote{it's similar to interacting with a young child, where you have to give very specific instructions and keep it concise} by P4 and \enquote{the experience felt somewhat like talking to young children, where you try to keep things simple in your communication} by P5. We believe this may have been amplified by the fact that the \ac{LLM}-based agents were happy to provide lengthy explanations and asked for clarifications whenever they were not sure what to do. This also happened in situations, where participants did not expect the \change{simulated robot agents} to have any problems understanding what they were told to do, e.g., P9 being surprised at the \change{agent}'s response when being asked to pick up an object it was currently not next to: \enquote{When the system mentioned that an object was out of its reach and asked for permission to approach it first, I found it unintuitive. I expected it to act on the instruction directly.}

\subsection{Dialog engagement vs. Command optimization}
\change{We observed two main ways in which the participants reacted to the simulated robot agents' answers. Some participants directly engaged with the dialogue, answering questions like \enquote{[...] Should I do that?}, which maintained the flow of conversation. On the other hand, another group of participants took a different approach: rather than continuing the dialogue, they sifted through the agents' responses for key details and crafted new, independent commands, streamlining the interaction.

For instance, when P6 and P9 asked Jupiter to place the vase next to the bed and Jupiter responded with \enquote{I can only place the vase on specific surfaces, like the bed or the ground. Can you specify where I should place it?}, P6 directly answered Jupiter's question with \enquote{on the ground}. In contrast, P9’s response, \enquote{Jupiter, please place the vase on the ground} was formulated as a complete command. Despite referencing elements from the ongoing dialogue, P9's instruction was structured in such a way that it could stand alone as an independent request without needing the context of Jupiter's prior question.}

Based on information from the interviews, a significant reason as to why participants did not respond to the \change{agents}' questions was a lack of trust in their ability to understand them. P14 alludes to this, while also mentioning the aspect of delay as a possible cause: \enquote{I perceived that they were offering a dialogue by asking, but I didn't believe they would understand me if I replied. I think it was also partly because, between their question and my answer, time had passed, making me think 'their memory was wiped'}. 
P6 also explicitly mentioned that their initial idea was to always say the names of the agents and the commands each time they spoke. This idea was reinforced, as the system's inability to decide the right recipient (which is very unlikely if the names are mentioned), can be a costly mistake.

\subsection{Holistic coordination vs. task decomposition}
We observed significant differences in the complexity of the participants' requests, especially when it comes to tasks that can be completed with multiple \change{simulated robot agents}. Some participants seemed to prefer breaking down tasks into simple steps, and wait for the \change{simulated robot agents} to complete one before moving on to the next step. On the other hand, some participants issued strategic multi-\change{agent} commands, giving each \change{agent} one or multiple steps to execute simultaneously. Generally, the participants who used more complex language seemed more inclined to try more elaborate instructions.

P11, for instance, issued complex commands like \enquote{first to you, Pluto. Put down the weight and fly to the refrigerator. To you, Neptune. Move in front of the refrigerator. To you, Jupiter. Pick up the weight after Pluto has put it down} and remarked in the interview that their initial impression that each \change{agent} had to be addressed individually had been subverted, and the instructions could be surprisingly elaborate and target multiple \change{agents} at once. Conversely, other participants broke down the tasks into the smallest possible steps and issued commands to each \change{agent} separately, such as P6 and P13. 

One reason for this seemed to be a lack of faith in the \change{agents}' abilities to understand longer commands, or a preconceived notion that the \change{agents} would perform better when given simple orders.
This coincides with the interviews, in which P6 likened their default mode of communicating to how they would talk to a dog and P13 mentioned that they tried to be \enquote{clear and concise} from the beginning. A similar interpretation was expressed by P4, who stated that  \enquote{I always had to think beforehand about what to say, to say it briefly and precisely, and yes, very clearly. I had to put myself in the system's shoes, guessing what it probably expected in the way I spoke.}
P4 was one of the participants who used very simple language from the beginning and intuitively broke down tasks into simple steps, likely expecting that the system would not be able to handle longer sequences directed at multiple \change{agents}. 
These patterns were exacerbated by the \ac{LLM}'s tendency to mirror the user's way of talking, responding more \enquote{robotically} when given \enquote{robotic} instructions. Additionally, in response to failure situations, it is not unreasonable to assume that participants would tone down the complexity of their commands. However, our data suggests that the participants generally decided on the complexity of their requests from the start and did not adjust its overall nature during the study, even in response to \change{agent} failure. 

\subsection{Overall aspects of User Experience}
In this section, we will present a collection of sub-themes describing the user's experiences with the system and their positive and negative impressions of it according to the interviews. These sub-themes range from the degree to which the system was lenient with input, to system feedback and responsiveness. At the end, we will present an overview of the \ac{SASSI} factors to corroborate these findings.

\subsubsection{Semantic flexibility}
\label{sec:semanticflex}
Participants appreciated that there was a high degree of flexibility to the words they were able to use. e.g., P14 said \enquote{I found the system's understanding impressive. I used various words like 'move', 'go there', 'fly there', 'proceed there', and I believe they were all understood. I didn't feel tied to a specific word, which made the interaction feel more natural.}
P3 echoed this statement, while also appreciating that the system did not have any issues with pauses during a request or exchanging pleasantries like \enquote{please} and \enquote{thank you}. 
It was expected that the \ac{LLM}, with its wealth of linguistic knowledge, would be fairly flexible when it comes understanding variations of words and commands. However, the results indicate that the \ac{LLM} was rather meticulous  when it came to words that are part of its function descriptions, sometimes refusing to do something if, for example, the name of an object was slightly off, as shown in the following exchange between P11 and Pluto:

\noindent\noindent\begin{tabular}{@{}p{2cm}p{15.3cm}@{}}
\texttt{{\small P11:}} & \texttt{{\small Pluto, please go to the blue trash can.}}\\
\texttt{{\small Pluto:}} & \texttt{{\small Sorry, I can only go to locations that are defined in my functions. A blue trash can is unfortunately not on my list. However, I can move to the trash bin, if that helps. Should I do that?}}\\
\end{tabular}

This is further substantiated by the interview data. P9 states that \enquote{they did execute the commands, but sometimes they were very meticulous}. While the example above is an outlier and the \change{simulated robot agents} could generally make that connection, it was apparent that they paid special attention to terms outlined by their functions. This phenomenon, and the fact that participants were not provided with any guidelines as to which words to choose, is also reflected in the \ac{SASSI} results. Question 29, which states \enquote{I sometimes wondered if I was using the right word}, received an average score of 5.92 (M = 5.92, SD = 1.73) out of 7, where 7 indicates 'strongly agree'.

\subsubsection{Verbal \change{agent} feedback}
The verbal feedback by the \change{simulated robot agents}  was generally regarded as helpful. Participants mentioned that the clarifications provided by the \change{simulated robot agents} made tasks easier and the \change{agents}' decisions more understandable. For example, P9 stated in their interview that \enquote{the robots behaved very understandably. Like with the key handover, Neptune couldn't directly take the key from Pluto but could pick it up from the ground. He explained that, so it was clear what I had to do.} 
In a similar vein, P6 called the feedback helpful and appreciated that the \change{agents} corrected their nomenclature in cases where P6 could not visually discern what type of object was in front of them. However, P6 also remarked that the feedback was too long for their taste, which aligns with statements made by P13 and P14. 

\subsubsection{Mismatch regarding expected robot autonomy}
\label{sec:robotautonomy}
Participants reported that they sometimes expected more autonomous behavior by the \change{simulated robot agents}, especially in situations that required a sequence of tasks to be executed. As it was up to the discretion of the non-deterministic \ac{LLM} to decide between calling a function or asking a user for clarification before proceeding, there was a certain amount of variance to the autonomy with which the \change{agents} executed given tasks. For example, P3 expressed: \enquote{In certain situations, I didn't realize I had to explicitly instruct the system to approach an object before picking it up. With humans, if I say 'pick up the key', I'd expect them to walk over to the key and then pick it up.}
This was referring to a situation in which P3 had asked Jupiter to pick up the key, to which Jupiter responded by asking for permission to approach the key first. P6 and P9 echoed this sentiment, expressing a similar surprise about the \change{agents} asking before executing a pre-requisite task. However, there were also occurrences where \change{an agent} would by itself decompose a task into necessary sub-tasks and perform them autonomously. For example, when P8 asked Jupiter to \enquote{take three plates and throw them in the trash bin}, the \change{agent} executed the correct function-triplet (consisting of movement, pick up, and throw away) three times in succession.

\subsubsection{Inherent response-lag}
A common complaint lodged by the participants was that the system took too long to answer. In the application, there was a visual cue to indicate the status of the transcription and which \change{of the simulated robot agents} were currently processing it to let the users know that they had been heard. However, there was an amount of delay before the \change{agents}' responses that was unlike an inter-human conversation, where feedback is usually instant. 
This led to cases in which participants started talking without waiting for the responses, as described by P9: \enquote{In terms of naturalness, the robots' feedback always came late. There was a lot of narration, and by the time they responded, I had already continued speaking}.

This is reflected by the \enquote{Speed} dimension of the \ac{SASSI}, where the system received an average score of 5.25 (M = 5.25, SD = 1.57) (see Table \ref{tab:sassi}), with 7 being the slowest. 

The reason for this behavior are due to the technical setup and we will discuss them in more detail in the discussion section.

\subsubsection{SASSI factors}
For an overview, \autoref{tab:sassi} shows the different dimensions of the \ac{SASSI} questionnaire, each on a scale from 1 to 7, averaged across the 12 participants with standard deviation.

\begin{table}[htbp]
    \centering
    \captionsetup{justification=justified}
    \caption{Descriptive statistics for individual \ac{SASSI} dimensions.
    \label{tab:sassi}}
    \small
    \begin{tabular}{lcccccc}
    \\
        \toprule
        \textbf{}	& \textbf{System Response Accuracy} & \textbf{Likeability} & \textbf{Cognitive Demand}  & \textbf{Annoyance} & \textbf{Habitability} & \textbf{Speed} \\
        \midrule
        $M$         & 4.43                              & 5.36                 & 4.88                       & 3.80               & 4.44                  & 5.25\\
        $SD$        & 0.95                              & 1.01                 & 1.02                       & 1.44               & 1.22                  & 1.57\\
        \bottomrule
    \end{tabular}
\end{table}

\change{In Table \ref{tab:sassi} we see that for dimensions accuracy, likeability, cognitive demand and annoyance our system is rated positively, while for the dimensions of speed and habitability it is rated on the negative spectrum of the scale. Due to the sample size of our study, the SASSI scores are difficult to interpret on their own. But we use them to contextualize our qualitative findings in the upcoming discussion section.}

\section{Discussion}
\change{Our findings suggest that some users did not perceive robots as equal conversation partners or at least modified the way in which they talked to the agents in a way that is not representative of a human-to-human interaction.}
Conversations between the participants and the \change{agents} often boiled down to simple instructions and some users effectively ignored the \change{agent}'s part in the conversation, despite the much more elaborate conversational capabilities of \ac{GPT}. \change{Agent} error and the overly meticulous nature of the \ac{LLM} in some cases could have been contributing factors to a user's low expectations of the \change{agent}'s abilities. We believe that, \change{in order} to allow for smooth cooperation, the \ac{LLM} must be aligned in such a way that its decisions are as understandable to the user as another human's. While our data shows attempts on the side of the \ac{LLM} to stay as aligned as possible by frequently communicating and asking questions, the \ac{LLM} was almost \enquote{too communicative}, leading to a mismatch regarding the expected \change{agent} autonomy. It is clear that a deeper, intuitive understanding is required on the \ac{LLM}'s side so that the \ac{LLM}'s and the human's conceptual models can be matched closely. 

\subsection{Lessons Learned}
\textbf{Function calls are a double-edged sword:} Function calls, with their reliable formatting and consistent nature, turned out to be very effective at connecting the language-based \ac{GPT}-model to coded \change{simulated} robot processes. Functions are highly modular and they are very adjustable, making them a good fit for an experimental framework that can hold multiple agents with vastly different abilities. This feature shows promise, even when applied to more elaborate scenarios, as function descriptions can be complex with multiple nested properties. The function feature also displayed the ability to confine the \ac{LLM} more effectively than a textual ontology approach. However, the feature created other issues in the model's behaviors. For instance, the cases in which the model was overly meticulous illustrated in \autoref{sec:semanticflex} were not conducive to intuitive cooperation. Likewise, the model's refusal to execute a function and ask the user for permission needlessly, as shown in \autoref{sec:robotautonomy} could have been a result of the function feature itself. Due to the opaque nature of \acp{LLM} it is difficult to say what exactly caused these particular problems. 
Overall, we believe that functions have proven to be an effective way for prototyping and testing \ac{GPT}'s abilities to make decisions that can be translated into \change{simulated robot} actions.

\textbf{GPT displays flaws as a solitary controller:} Our results suggest that \ac{GPT} as a solitary controller for an actual robot is not yet feasible, especially when it comes to usability concerns.
Perhaps the most glaring issue of \ac{GPT} as a solitary controller is its inherently non-deterministic nature. Even when the model is given a low temperature (a parameter which controls the randomness and creativity of the responses), the outputs are too inconsistent to act as a governing element for a real robot. 

\change{Another apparent issue is \ac{GPT}'s inability to generate actionable plans on its own. While our study has shown instances of the models being able to decompose complex tasks successfully, in most cases it will be unable to piece together what tasks have to be executed in which order to fulfill an overarching goal. Unless the model improves to a point where it can overcome these issues on its own, additional modules would have to be developed to combat these problems. A potential approach within this framework could involve diversifying the agent's solitary \ac{LLM} instances into multi-agent systems, where multiple \ac{LLM}s assume specialized roles, supplemented by modules like a dedicated planner to mitigate these limitations. 
Incorporating a dedicated module for semantic disambiguation and tracking dialogue states could also enhance the already human-like conversations facilitated by the \ac{LLM} and improve the user's perception of interacting with the agents.

A more sophisticated way of formalizing the task domain and communicating this formalization properly to the model could also enhance performance. In this context, \ac{GPT}'s function capabilities seem suitable and could be further improved by introducing probabilities and affordance functions, similar to Google's \enquote{SayCan} \citep{ahn2022can}. In this way, a task could be broken down into a set of variables, which is represented by a range of functions and auxiliary information. A function's likelihood to succeed (based on  the agent's current state, task dependencies, task structure and other variables) can then be communicated to the model and allow it to make more informed decisions.

Another flaw of \ac{GPT} as a solitary controller is tied to one of \ac{GPT}'s greatest advantages: its massive parameter count. While it endows the model with a wealth of linguistic knowledge, it also limits its deployment to a physical robot. Being a cloud-based service, the \ac{GPT} \ac{API} introduces latency which is exacerbated by problems with the internet connection. Introducing additional \acp{LLM} in a future multi-agent approach would only further compound this issue.}
The system's slowness is reflected within the \ac{SASSI}'s factor 7, as well as the statements made by the participants during the interviews. Aside from negatively impacting the user experience, instant feedback is essential to language-based communication, especially during tasks where time-critical cooperation is required.

\change{\textbf{Calibration could be very beneficial:} We believe that a calibration process, in which a user and the \ac{LLM} clarify what their preferred way of communicating is, could be helpful. Our findings indicate a general assumption among humans that their interaction with robots lacks the dimension of meta-communication. Implementing a calibration phase not only has the potential to align initial expectations but also to mitigate issues such as overly detailed explanations that may hinder the efficiency of communication. 

Another promising avenue would be the introduction of additional communication channels, like an empathetic channel, specifically designed to allow a user and the robot's \ac{LLM} to exchange feelings and emotions with one other. Such a channel could improve the explainability of an agent and further strengthen its alignment regarding a user's individual preferences. To this end, employing a theory of mind approach similar to \citeauthor{scassellati2002theory} could refine the agent's contextual awareness and feedback mechanisms, making it easier for the user and the agents to empathize with each another and improve communication.

However, introducing emotional exchanges in human-agent interactions, such as through an empathetic channel, poses ethical questions, especially regarding the authenticity of the agents' emotions. The concern centers on maintaining genuine interactions and preventing user deception. Transparency about the agents' capabilities and ensuring their beneficial use is important and as such, ethical insights and guidelines provided in works by \citet{breazeal2003emotion} and \citet{lin2014robot} should be at the forefront when considering such implementations.}

\textbf{A competent sensory system for the robots is essential:} During communication, certain assumptions arise about what the conversation partner knows, sees and feels. What we have seen in our study is the importance of the \ac{LLM}'s conceptual model to be aligned to the user's. For one, this will necessitate the inclusion of non-verbal communication, such as pointing. Additionally, an improved sensory system would allow for more varied and flexible interpretation of user requests. The more data the \ac{LLM} has access to when deciding whether to execute an action, the more informed its decision will be. Supplying the \ac{LLM} with sensor and camera information from a physical robot therefore shows a lot of potential but will introduce a new set of challenges.

\change{As an addendum, at the time of our study, \ac{GPT} was yet unable to take images as input. In the future, incorporating this functionality into our framework would provide \ac{GPT} with another way by which it can ascertain the state of the virtual world. This would make the framework more robust, as the visual information would supplement the textual descriptions of the scene and provide the model with a way to check the virtual world "for itself". This could potentially even resolve cases where the model refuses to believe information given to it through system messages, as it would gain the ability to take a picture of the scene and receive additional information about the actual state of the virtual world.}

\textbf{Additional avenues for control are necessary:}
Intervention is very important when interacting with a robot that can act autonomously and perhaps even more so when interacting with multiple. Our study indicated an additional need for intervention as well. The issue with intervention in this context is that, due to the processing time of the transcription and \ac{GPT} modules, the regular channels of communication between the user and the \change{simulated robot agents} are unfit for intervening. Therefore, additional possibilities of intervention will have to be explored. An intermediary solution would be the introduction of a button that immediately stops the \change{simulated robot agent} regardless of the \ac{LLM}'s current state. 
Another addition to improve the controllability of the system would be to make the processes of the \ac{LLM} less opaque and find ways to visualize its current state. This will be especially important as task complexity and robot complexity grow.

\textbf{Inter-robot communication is needed:}
While the framework shows promise in allowing a user to interact with multiple \change{simulated robot agents} simultaneously in an unstructured way, it is clear now that additional communication between the \change{agents} themselves is required. While the \change{agents} share a general sense of what their fellow \change{agents} are doing, textual representations and a unified log of conversations is insufficient, especially during tasks where the \change{agents} have to collaborate. Their responses should also be clustered in a way that the response to the user is streamlined and free of repetitions. 

The nature of our framework would support such approaches, with the agent controller potentially taking on a more elaborate part in controlling agent-to-agent communication and interaction. The agents themselves could even exchange function calls with each other, allowing for synchronized and planned cooperation towards certain goals.

\subsection{Limitations}
\label{sec:limitations}
There are a few noteworthy \change{limitations} that pertain to the simulation framework, as well as the study the framework was used in.

\change{First and foremost, our study results were obtained in a simulation environment in \ac{VR} with simulated robot agents. Therefore, it is difficult to say to what extent results may transfer to physical robots. Most importantly, the simulation simplified any potential problems from physical actions, such as a robot slightly missing an object while trying to grab it. However, as we see the benefits of language communication for variable autonomy more on the tactical, strategical and cooperational layer, this was a conscious decision to reduce complexity and allow us to focus on these areas.}

Utilizing an \ac{LLM} comes with certain \ac{LLM}-inherent limitations. First off, due to their opaque nature, programs utilizing \acp{LLM} are notoriously hard to debug. While there were some unforeseen consequences caused by unexpected \ac{LLM} behavior, resulting bugs during the study only caused minor inconveniences and all participants were able to solve the tasks successfully. The use of German language could have been a possible cause for some of these unexpected behaviors, especially in regards to functions - although the objects described in the function descriptions were deliberately given German names to prevent the \ac{LLM} from disambiguating.
Additionally, the framework inherits other \ac{LLM}-related limitations, such as possible biases in training data and errors being propagated forward, as false output becomes part of future input.

Another limitation arose from the University-based internet connection used for the study. While the cloud services employed in this application inherently involve some latency, the sometimes unreliable internet connection further compounded the issue. While there were no disconnects or complete outages, the delays in transcriptions and model responses were at times noticeably longer, in contrast to quicker performance observed with a stable internet connection.

\change{Additionally, while we endeavored to design the tasks in a way to avoid undue influence on participants' responses, it is important to recognize that a certain degree of influence from the task characteristics is challenging to completely avoid. Therefore, another limitation that should be mentioned is the influence the task design had on the perception of the users. As mentioned previously, in an attempt to avoid biases from pre-defined task descriptions, we included a virtual task goal screen that only showed the desired goal state. Based on this preview, users had to infer what commands to give to the agents based on that. For the more difficult tasks, such as task 5 and 7, this information may have been insufficient for some participants. As a result, these tasks were not as clear-cut as the remaining ones and generally required more trial and error, which may have resulted in more negative experiences with the agents in these particular tasks.}

\section{Conclusion}
\change{This study makes a significant contribution to understanding the integration of \acp{LLM} such as \ac{GPT} in a simulated human-robot teaming environment and the interaction dynamics between a user and \ac{LLM}-based agents when they have to communicate and collaborate on a task.} Our findings indicate that despite the advanced capabilities of \acp{LLM}, user perception and interaction with \change{simulated robot agents} present challenges. The study highlights that users do not perceive robots as equal conversational partners, leading to a communication limited to simple instructions. Still, the study also showcased more elaborate conversation for those users who did not expect preconceived limitations in the conversational abilities of their \change{simulated robot cooperators}.

A critical aspect is the alignment of the \ac{LLM} to make decisions understandable and predictable for the user. The over-communication by the \ac{LLM} and the resulting ambiguity about \change{agent} autonomy demonstrate the need for a more balanced approach. Furthermore, the study underscores the importance of a calibration process to align communication preferences between users and \ac{LLM}.

The study also reveals that using \ac{GPT} as a sole controller for \change{simulated robot agents} has limitations, primarily due to its non-deterministic nature and latency issues from cloud connectivity. This underscores the necessity of considering \acp{LLM} as part of a broader system that incorporates both user feedback and sensor information.

In conclusion, our research provides valuable insights into the dynamics of human-robot teaming, indicating a significant need for further research and development to effectively deploy \acp{LLM} in such systems. Our findings suggest that a multidisciplinary approach, encompassing technology, user experience, and psychological aspects of interaction, is required to fully realize the potentials of \acp{LLM} for variable autonomy in human-robot teaming.

\bibliographystyle{Frontiers-Harvard} 
\bibliography{MainPaper}

\end{document}